\documentclass[conference]{IEEEtran}

\usepackage{tabularx}
\usepackage{caption}
\usepackage{cite}
\usepackage{amsmath,amssymb,amsfonts}
\usepackage{graphicx}
\usepackage{textcomp}

\usepackage[hyphens]{url}

\usepackage{verbatim}
\usepackage{amsmath,amsfonts,fancyhdr,graphicx,textcomp,listings,multirow}
\usepackage{url}

\usepackage{xcolor}
\usepackage[hidelinks, urlcolor = blue]{hyperref}
\usepackage{algorithm}
\usepackage{arevmath}     % For math symbols
\usepackage[noend]{algpseudocode}

\definecolor{pink}{cmyk}{0, 0.7808, 0.4429, 0.1412}
\definecolor{codegreen}{rgb}{0,0.6,0}
\definecolor{codegray}{rgb}{0.5,0.5,0.5}
\definecolor{codepurple}{rgb}{0.58,0,0.82}
\definecolor{backcolour}{rgb}{0.95,0.95,0.92}
\newcommand{\amc}{AMC}
\newcommand{\amcCache}{AMC Cache}

\def\BibTeX{{\rm B\kern-.05em{\sc i\kern-.025em b}\kern-.08em
    T\kern-.1667em\lower.7ex\hbox{E}\kern-.125emX}}

\title{AMC: Access to Miss Correlation Prefetcher for Evolving Graph Analytics}
\author{\normalsize{Abhishek Singh\IEEEauthorrefmark{1}\IEEEauthorrefmark{2}, Christian Schulte\IEEEauthorrefmark{3}, Xiaochen Guo\IEEEauthorrefmark{1}\IEEEauthorrefmark{2}}\\
 
 \IEEEauthorblockA{\IEEEauthorrefmark{1}Lehigh University \IEEEauthorrefmark{2}Samsung Semiconductor, Inc., USA \IEEEauthorrefmark{2} Columbia University\\
    }}
\begin{document}
\maketitle
\thispagestyle{plain}
\pagestyle{plain}

%%%%%% -- PAPER CONTENT STARTS-- %%%%%%%%

\begin{abstract}
Modern memory hierarchies work well with applications that have good spatial locality. Evolving (dynamic) graphs are important applications widely used to model graphs and networks with edge and vertex changes. They exhibit irregular memory access patterns and suffer from a high miss ratio and long miss penalty.  Prefetching can be employed to predict and fetch future demand misses. However, current hardware prefetchers can not efficiently predict for applications with irregular memory accesses.
 
In evolving graph applications, vertices that do not change during graph changes exhibit the same access correlation patterns. Current temporal prefetchers use one-to-one or one-to-many correlation to exploit these patterns. Similar patterns are recorded in the same entry, which causes aliasing and can lead to poor prefetch accuracy and coverage. 
%With programmer\textquotesingle s knowledge, one can identify the vertex and its dependents traversal to create `many-to-many' correlation to resolve the aliasing problem. 
This work proposes a software-assisted hardware prefetcher for evolving graphs. The key idea is to record the correlations between a sequence of vertex accesses and the following misses and then prefetch when the same vertex access sequence occurs in the future. The proposed Access-to-Miss Correlation (\amc{}) prefetcher provides a lightweight programming interface to identify the data structures of interest and sets the iteration boundary to update the correlation table. For the evaluated applications, \amc{} achieves a geomean speedup of 1.5$\times$ as compared to the best-performing prefetcher in prior work (VLDP). \amc{} can achieve an average of 62\% accuracy and coverage, whereas VLDP has an accuracy of 31\% and coverage of 23\%.

\end{abstract}

\section{INTRODUCTION}

Single-thread performance is vital to system performance improvement. 
Data prefetching is a proven technique to improve single-thread performance by overlapping the miss penalty gap with computation to hide long miss latency. A prefetcher predicts and fetches a cache line before its demand miss in faster memory (L1D, L2C) from slower memory (LLC, main memory). This helps to reduce pipeline stalls on waiting for memory. Over the past few decades, numerous prefetching mechanisms have been proposed targeting on different types of memory access patterns. The key differences reside in the types of exploited correlations (e.g., PC-address \cite{bingo,t2,dol}, PC-offset~\cite{bestoffset,bingo}, address-address~\cite{markov, vldp}, etc.) and the types of targeted patterns (e.g., stride~\cite{t2,dol}, stream~\cite{stream}, irregular~\cite{misb,isb,triage,vldp}).

Evolving graphs (a.k.a. dynamic graphs)~\cite{Ogdynamic,dynamicGraph,graphone,tdgraph,Kineograph,jetstream,dredge,graphbolt} are the graphs that change over time. The two types of graph dynamics are \textit{vertex dynamic}, wherein the vertices set changes during computations, and \textit{edge dynamic}, wherein the edges are added and deleted from time to time. Many important applications~\cite{timeDyGraph,streamgraph,telegraphcq,dynamicGraph} use dynamic graphs to model complex relationships that change over time, such as recommendation systems~\cite{tang2021dynamic}, internet of things~\cite{wen2017fog}, and social networks~\cite{twitter}.
%In modern processor~\cite{intel_pref}, multiple prefetchers have complimentary heuristics targeting different access patterns (for strided/sequential memory accesses). However, these prefetchers still need help to crack a heuristic to help predict irregular memory accesses in graph applications~\cite{ligra}. 

\begin{figure}[ht]
\centering
\includegraphics[width=1\linewidth]{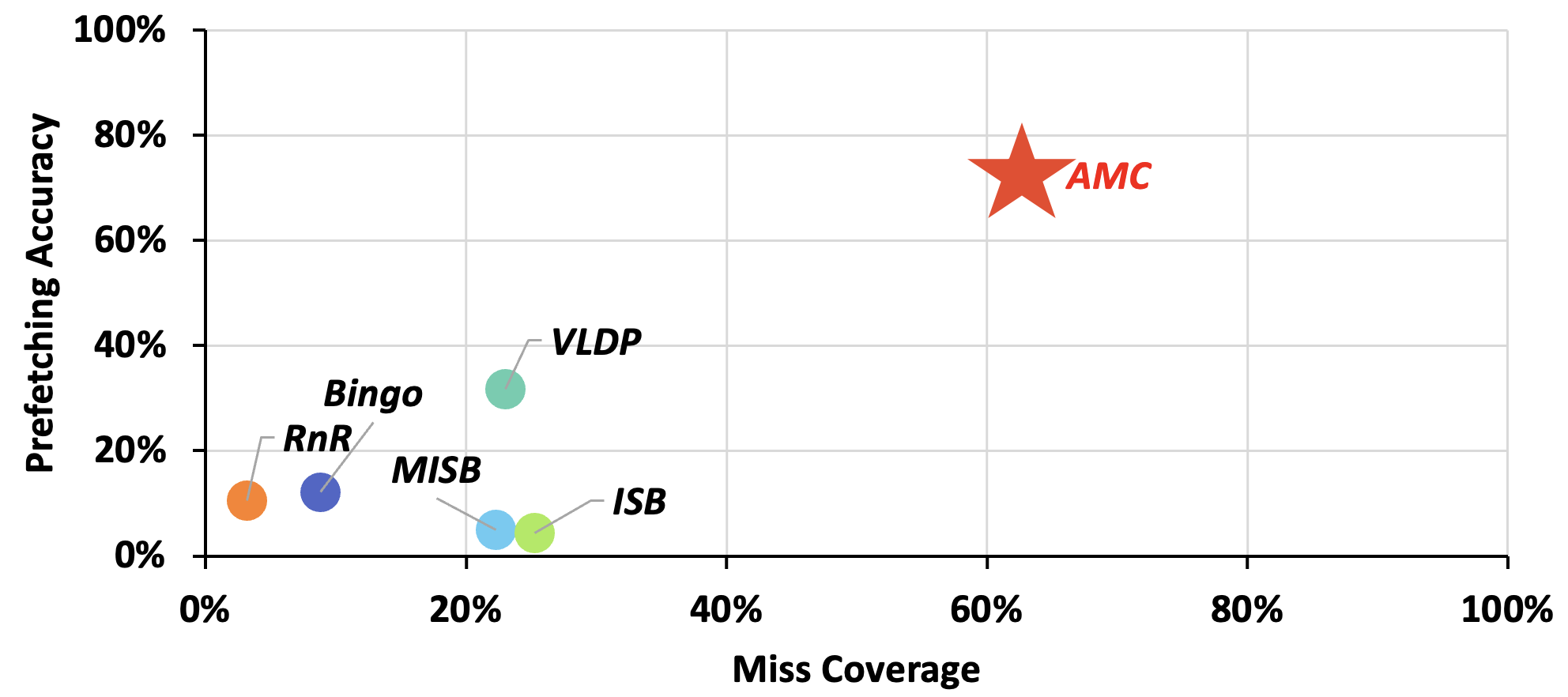}
  % \vspace*{-8mm}
\caption{Prefetcher coverage and accuracy of PageRankDelta \cite{ligra} on amazon \cite{datasets} graph}
   \vspace*{-2mm}
  \label{fig:cov_acc}
\end{figure}

Existing hardware prefetchers~\cite{misb,isb,markov,sms,stms,rnr,triage,vldp} have tried to exploit repeating patterns on different correlations. However, due to the lack of contextual information in building these correlations, existing prefetchers achieve limited performance improvement for dynamic graph applications. 
%Recent irregular hardware prefetchers~\cite{isb,misb,triage,bingo,vldp} fail to adapt to the changes of the input graphs because they are not using enough contextual information~\cite{thesisMisb} in developing correlations in access patterns. 
On the other side of the spectrum, software prefetchers rely on programmer\textquotesingle s expertise to issue prefetch instructions for future needs. This may help to fetch accurately but can increase instruction count, which might cancel out the performance gain. Additionally,  software prefetchers have less or no knowledge of run-time system dynamics (\textit{e.g.}, control flow, bandwidth utilization, cache conflict), which makes timely and accurate prefetching very challenging.
Recently proposed hardware-software cooperative prefetchers ~\cite{prodigy,metasys,droplet} fully utilize the inherent relationship between data structures in \textit{static} graphs. These prefetchers use sequential dependencies for prefetching different data structures in the graphs, which can lead to late prefetchers. For example, DROPLET~\cite{droplet} triggers vertex data prefetching only when DRAM services edge miss, which is often too late~\cite{rnr}.  

An ideal irregular prefetcher should be (1) able to prefetch \underline{accurate} irregular data and strategically place it in caches that strike a balance between cache contention and coverage.
 (2) able to perform \underline{timely} prefetch, \textit{i.e.}, adapt to application phases that changes. 
 (3) able to \underline{cover misses} that stalls processor. 
 Vertex property array access in graph analytics is responsible for most misses due to indirections used in graph data structures~\cite{gretch}.
 The vertex-neighbor relationship in graph analytics typically remains intact even in the dynamic graphs, when vertex/edges are added/deleted at run time. One can exploit this relationship to develop a correlation between vertex accesses and misses on other data structures, thus adding contextual information to the correlation.
This paper proposes a novel software-assisted \textit{``Access-to-Miss Correlation (\amc{}) hardware Prefetcher''}  to issue \textit{accurate} and \textit{timely} prefetches at the L2 cache.
\amc{} selects L2 cache as its prefetch destination. This is based on the observation in DROPLET~\cite{droplet} that using the L2 cache for prefetching leads to negligible cache pollution for graph applications.
\amc{} uses L1 cache word accesses and L2 misses to form fine grain access-to-miss correlation.
Using lower-level cache access as a trigger to prefetch higher-level cache misses provides good prefetch timeliness.
A lightweight programming model (Section~\ref{sec:softwareSupport}) allows the programmer to choose a data structure (\textit{e.g.}, vertex array) as the target data structure.
The \amc{} prefetcher records the cache misses in between target data structure accesses to create access-to-miss correlation entries and updates these entries at run time. 
In addition, \amc{} exploits the existing one-to-one correlation between the graph data structures (frontier-vertex array) as well. %In dynamic graph applications, frontier array is responsible to hold active vertices in the current iteration~\cite{ligra}.

Fig \ref{fig:cov_acc} shows the prefetching accuracy and L2 miss coverage comparison of the proposed \amc{} prefetcher with five existing prefetchers when running a PageRankDelta (PGD) application~\cite{ligra}: two spatial prefetchers (Bingo~\cite{bingo}, VLDP~\cite{vldp}, and three temporal prefetchers (MISB~\cite{misb}, ISB~\cite{isb}, and RnR~\cite{rnr}).
The key innovations of the proposed ~\amc{} prefetchers are: 
\begin{itemize}
    \item
    \textbf{\amc{} uses a lightweight programming model to record access-to-miss correlations for prefetching in dynamic graphs}.
    Previous hardware-software cooperative prefetchers~\cite{prodigy,droplet,metasys,rnr} either record the miss sequence directly or rely on programmer/compilation technique  to analyze detailed graph data structure dependency.  
    %Existing hardware prefetchers use only memory access patterns to develop correlations~\cite{isb,misb, bingo,vldp,triage}.
     The proposed \amc{}\textquotesingle s lightweight interface only require programmer to identify only two data structures and uses underlying hardware to develop an access-to-miss correlation for prefetching, which adapts to the changing nature of dynamic graphs.
    \item
    \textbf{\amc{} exploits a novel \textbf{many-to-many correlation} between target data structure accesses and other data misses}. Existing prefetchers ~\cite{markov,misb,isb,bingo,sms,vldp,triage} use one-to-one, one-to-many, or many-to-one correlations, which cannot distinguish similar memory access patterns and can lead to inaccurate prefetches~\cite{thesisMisb}. \amc{} uses a sequence of target accesses as the triggering event to provide contextual information to distinguish similar memory access patterns accurately.
    \item
    \textbf{\amc{} uses an on-chip SRAM to cache miss stream in FIFO order and compress the miss stream to reduce off-chip traffic and storage}. Prior works~\cite{ghb,misb,isb,triage,vldp,markov} either used a tabular or associative cache to store misses, which leads to a sizeable on-chip area to store metadata. \amc{} stream metadata in FIFO order to simplify the on-chip storage (Table~\ref{tab:prefetcherConfig}) and use Base$\Delta$ compression~\cite{baseDeltaCom} to reduce off-chip traffic and storage.
   
\end{itemize}
\section{BACKGROUND AND RELATED WORKS}

\begin{table*}[ht]
\footnotesize
% \begin{small}
\begin{tabularx}{\textwidth} { 
  | >{\centering\arraybackslash}X 
  | >{\centering\arraybackslash}X 
  | >{\centering\arraybackslash}X 
  | >{\centering\arraybackslash}X 
  | >{\centering\arraybackslash}X |}
  
 \hline
 \textbf{Prefetcher Design} & \textbf{Correlation Style} & \textbf{What to Prefetch} & \textbf{Storage Format
} & \textbf{When to Prefetch}\\
 \hline
 \textbf{AMC} (Proposed) & many-to-many: target access stream - miss addr stream  & misses other than the target data structure & compressed miss stream  & target addr access \\

 \hline
 \textbf{RnR~\cite{rnr}}  & one-to-many: window count - offset stream & defined by software & Irregular data structure offset & software assist, replay timing control mechanism \\

 \hline
 \textbf{ISB~\cite{isb}}  & one-to-many: PC - addr stream  & No constraint & TLB dependent compressed format  & cache access \\

 \hline
 \textbf{MISB~\cite{misb}}  & one-to-many: PC - addr stream  & No constraint & 8-byte (single mapping)  & cache access \\
 
 \hline
\textbf{Bingo~\cite{bingo}}  & one-to-many: PC - addr/offset stream, addr - addr stream & No constraint & on-chip tage-like history table  & cache miss in a new page \\

 \hline
 \textbf{VLDP~\cite{vldp}} & one/many-to-one: delta/page offset - delta & No constraint  & cascaded recent delta table & cache access\\

\hline
\end{tabularx}
\caption{Comparison to other prefetchers}
  \vspace*{-4mm}
\label{tab:pref_comp_design}
% \end{small}
\end{table*}
 This section discusses the background of evolving graph applications, prior work on prefetchers, and recently proposed accelerators for graph applications. 
%\amc{} prefetcher is compared with HW-SW cooperative, temporal, spatial, pointer-based, and software prefetchers. Then, dynamic graph accelerators are discussed.
%Table~\ref{tab:pref_comp_design} compares the key differences between the proposed \amc{} prefetcher and others.

\subsection{Evolving Graphs}
%TODO: Expend and rewrite this subsection. Talk about the two types of evolving graph applications considered in this work. Common features of these applications that allow AMC to work well (frontiers, indirections, etc.). Or what are the important assumptions made about these applications. 

Applications that use graph-based algorithms and data structures in real-world scenarios where the relationship constantly evolves between entities are known as evolving graph applications. This work uses two types of evolving graph application: iterative graph algorithm with early convergence and graph applications with changing input graphs. 
Early convergence iterative graph algorithms, like PGD~\cite{ligra} offer an optimization over PageRank~\cite{page1999pagerank}. %While PageRank requires 100s~\cite{rnr} of iterations to compute page rank values for a given input graph, PGD requires fewer than 50\%  of the PageRank iteration to calculate approximate page rank values. 
PGD typically requires fewer iterations as compared to PageRank and has a faster runtime. 
This is possible because PGD only updates vertices in an iteration whose PageRank value has changed by more than some $\delta$-fraction. Therefore, in every iteration, a set of active vertices are involved in the PageRank calculation, resulting in less computation but non-repetitive irregular memory access patterns.
%This work considers early convergence iterative applications as a subset of evolving graph applications wherein vertex set changes with time~\cite{Ogdynamic}. 
Section~\ref{sec:MOTIVATION} discusses performance challenges with such irregular patterns. For graph applications with changing input, the method explained in Section~\ref{sec:setup} is used as the inputs to dynamic graph applications, which is similar to prior work~\cite{tdgraph,graphbolt}. 

Evolving graph applications typically use a frontier array, which is a bit map, to keep track of the vertices participating in the upcoming iteration or computation. This establishes a one-to-one correlation between the frontier and vertex accesses. Additionally, the inherent vertex-to-neighbor correlation is a one-to-many correlation between the vertex and its neighbor accesses. This can be used to fetch data structures related to the vertex present in the frontier. %These two assumptions hold true in critical real-world analytics applications.
\amc{} take advantage of these two properties of evolving graphs to build the correlations explained in Section~\ref{sec:MOTIVATION}. %In summary, \amc{} uses frontier array access stream to prefetch the \amc{} metadata entries on-chip to prefetch data structures before its demand related to the demand vertex access stream.

\subsection{Prior Work on Prefetchers}

AMC is a hardware-software cooperative prefetcher. The closest related works in the same category are  RnR \cite{rnr}, DROPLET~\cite{droplet}, and Prodigy~\cite{prodigy}.
%co-design unlocks cross-layer optimization by conveying application-specific knowledge to the hardware. %This information, also known as metadata, enriches the data with information that makes it easier to find, use and manage at the system level. 
%Prefetchers~\cite{rnr,prodigy,metasys,droplet} uses the programmer's knowledge to mark data structures and communicate the dependency among them to the system. 

DROPLET~\cite{droplet} uses a specialized malloc function to identify a graph application's targeted data structure (vertex and vertex property). DROPLET generates the addresses for an indirectly accessed vertex property value by prefetching the edge array. 
DROPLET~\cite{droplet} triggers vertex data prefetching only when DRAM services edge, which is often too late~\cite{rnr}.
Prodigy~\cite{prodigy} uses either compiler profiling or program annotation to generate data flow graphs of graph data structures. It uses demand access to the vertex node to prefetch the next vertex node and waits for the vertex node to be filled at the destination cache to initiate prefetch for its outgoing edges using the prefetched data. This requires a complete software stack change, including rewriting code, compiler, and OS just to optimize the prefetching of graph data structures.
The RnR prefetcher~\cite{rnr} targets on long, repetitive, irregular memory access patterns in iterative algorithms. It improves cache miss coverage and accuracy by recording in the initial iteration and replaying the miss patterns for prefetching in the following iterations. In dynamic graphs, wherein the vertex/edges change over time, RnR does not work well. AMC solves this issue by recording the access-to-miss correlations that are preserved in dynamic graphs.

%It requires TLB access to initiate prefetching for edges.
%In modern systems, with multi-level TLB, the TLB access latency may affect the prodigy's prefetching timeliness, this can be overcome by employing a high increase in the look-ahead distance ~\cite{event_driven} discussed later
%In modern systems, with multi-level TLB, the TLB access latency may affect prodigy's prefetching timeliness due to .
 
%MetaSys~\cite{metasys} is an open-source full-system FPGA-based infrastructure to evaluate various cross-layer optimizations in real hardware. It presents the HW-SW cooperative prefetcher as a case study. The software interface marks data-dependent data structures in graphs. On access to the trigger data structure, the prefetcher looks ahead and prefetches contents of the vertex, edge, and property array based on the computed index.
Similar to Prodigy, a case study on a HW-SW cooperative prefetcher presented in MetaSys~\cite{metasys} also rely on sequential dependency between data structures. 
%Both Prodigy~\cite{prodigy} and MetaSys~\cite{metasys} develop a sequential dependency between data structures in an application similar to DROPLET. 
For dynamic graphs, another limitation of these prefetcher
{\color{red}~\cite{prodigy,metasys,event_driven}} is their inability adapt to runtime dynamics (conditional branch). For example, PGD avoids redundant computation by examining only the vertex whose page rank value changed by a set threshold in the previous iteration. These prefetcher fails to account for control-flow knowledge for prefetching.
\amc{} overcomes the dependency challenge by using a single data structure as the triggering data structure to prefetch all of the other misses. In order to adapt to vertex and edge changes in the dynamic graphs, \amc{} continuously updates the correlations in every iteration and uses the latest one to prefetch.

%\subsection{Temporal Prefetchers}
\label{sec:temporalPrefs}
Temporal prefetchers~\cite{triage,misb,markov,isb} record memory access and then correlate it to either its PC or the previous access. %It works better for pointer chasing applications, wherein the application has static input.
They typically have high metadata storage overhead because they store a long sequence of memory addresses and inability to delete useless metadata. %Moreover, the history table size grows with the applications' memory footprint.
The closest related works to AMC in this category are ISB~\cite{isb} and MISB~\cite{misb}. 
ISB~\cite{isb} uses TLB and structural addresses to map physical memory addresses to structural addresses and store them using PC localization. It suffers from high metadata overhead, does not scale with large page sizes, and does not work with modern hierarchical TLBs. 
MISB~\cite{misb} solves this problem by employing the next-line prefetcher for metadata access as the structural address space is spatial and removing the TLB dependency to manage metadata caches. 
Unfortunately, when application\textquotesingle s input size grows, the metadata also grows. The problem is more severe for on-chip only prefetchers like Triage~\cite{triage} because they do not have off-chip metadata to fall back on to record growing metadata.
\amc{} prefetcher continuously updates the correlation table with only latest ones and compresses the metadata to reduce the storage overhead.
Temporal prefetchers~\cite{isb,misb,triage,markov,optMarkov} also suffer from aliasing problem~\cite{thesisMisb}. Multiple addresses can correlate to the same trigger event, which causes aliasing. \amc{} prefetcher solves this problem by linking multiple target accesses with miss stream, thus adding contextual information to correlation (Section~\ref{sec:MOTIVATION}). 
DVR~\cite{dvr} is a recently proposed architecture over VR~\cite{vr} targeting C[hash(B[hash(A[i ])])].
DVR utilizes vector functional units and vector registers to execute indirect memory instructions in advance. It groups together instructions with the same offset to a single vector instruction. However, DVR's progress in extracting MLP can be impeded if a previous instruction encounters branch misprediction, and it heavily relies on core structures such as ROB, VRAT, and stride detectors. An evolving graph uses a conditional branch instead of a hash function and may not utilize vector functional units effectively. However, it may be less effective in scenarios where the number of iterations is low, such as in BFS, because the DVR points themselves. This is according to a study by~\cite{dvr}.
In contrast, \amc{} is decoupled from the core's microarchitecture components. It records and intelligently replays indirect memory data structure to improve performance. It depends on the previous iteration recording to extract MLP and fully utilize L2's MSHR without competing demand loads.

%\subsection{Spatial Prefetchers}
Spatial prefetchers ~\cite{vldp,bingo,sms,stream} exploit the address delta similarity between cache accesses among different memory regions, which arise due to a fixed and regular memory layout of data objects. Such memory address patterns are common in server applications~\cite{oltp} (e.g., OLTP, DSS). The advantage of such spatial prefetchers is that they require less metadata.
VLDP~\cite{vldp} targets irregular access patterns within a page. Unlike a regular access pattern prefetcher, VLDP tries to predict a common pattern amongst past deltas. It uses TAGE-like table~\cite{tage} to solve the aliasing, leading to better accuracy. 
TAGE-like history refers to using multiple history lengths stored in various tables. In this approach, the prefetcher looks up multiple history tables to generate predictions rather than depending on a single history table for predicting future memory access patterns. 
It considers multiple history lengths and offers better prediction accuracy than a single history table. Moreover, it can adapt to changing application phases and capture complex correlations between memory access patterns, making it applicable to dynamic graph applications. Finally, it lowers the aliasing probability by utilizing multiple history tables.
BINGO\cite{bingo} also uses an optimized Tage-like table wherein the multiple history tables are fused into a single unified table and looked up multiple times with different history lengths. This reduces the overall storage overhead of traditional Tage-like predictors.
\amc{} also uses multiple trigger accesses. The novelty of \amc{} is to use accesses of only the targeted data structure as the trigger, which helps to improve the prefetching accuracy for evolving graph applications.
 
Software-based prefetching~\cite{compiler2003, software_prefetching, ainsworth2017software} for linked data structures requires programmer/compiler analysis to identify pointer-chasing access responsible for cache misses. This often requires significant effort to generate effective prefetch requests well ahead of demand requests to generate timely prefetches. Ainsworth and Jones proposed~\cite{ics} a configurable prefetcher aimed at improving performance for graph workloads. However, it only targets specific traversals for a certain graph format. Event-triggered programmable prefetchers~\cite{event_driven} employs an array of mini programmable prefetcher units to target heterogeneous access patterns using compiler profiling and maximize memory level parallelism, particularly in A[B[C[i]]], wherein array C can be prefetched in before its demand, which can lead to the prefetching of arrays B and C. ATP\cite{informedPref} explains the hardware complexity of Ainsworth and Jones's proposed prefetcher for indirect memory access and has timeliness problem similar to DROPLET~\cite{droplet}. ATP uses instructions to communicate data structure knowledge and a similar strategy as IMP to calculate linked data structures. Additionally, \amc{} does not rely on data-based prefetching, but instead relies on its previous recordings for prefetching.
These compiler profiling~\cite{compiler2003, software_prefetching, ainsworth2017software,ics,event_driven, apt_get} require software stack change, including rewriting the code, updating the compiler, and changing the operating system to optimize the prefetching of data structures. Additionally, they do not adapt well to run-time changes due to context switches or speculation misprediction. Furthermore, software prefetching increases the overall instruction code size.

Table~\ref{tab:pref_comp_design} summarizes the key differences between \amc{} and its closely related prefetchers.
%\subsection{Pointer-Based Prefetchers}
%Pointer-based prefetchers~\cite{imp,stream_chain,jump_pointer} target the a[b[i]] indirect accesses. It uses hardware and software/compiler methods to predict the address by the pointer. Ill-timed prefetching is a common problem in such prefetchers. IMP~\cite{imp} issues prefetch by predicting the correlated index stream. It suffers from low coverage and low accuracy due to bad timing. Using PC localization, stream chaining~\cite{stream_chain} prefetchers tries to overcome the timing problem. Compiler support prefetcher~\cite{event_driven} focuses on predicting c[b[a[i]]]. It exploits the data structure knowledge, i.e., array `a' shows MLP, and prefetching `b' and `c' need to immediately start the next demand access to array A.

%\subsection{Software Prefetchers}
%Software prefetchers~\cite{software_prefetching,apt_get} insert prefetch instructions at compile time and use performance counter~\cite{perform_counter} to identify critical performance loads, the perfect site to place prefetch instructions and the prefetch degree.  They do not adapt well to run-time changes due to context switches or speculation misprediction. Furthermore, software prefetching increases the overall instruction code size.

\subsection{Dynamic graph accelerators}
Accelerators for dynamic graphs have been proposed \cite{tdgraph,Kineograph,jetstream,dredge,hau,grasu,minnow} as stand-alone accelerators or near/in-memory processing engines. These accelerators often necessitate custom hardware design and programming models. \amc{} leverages the existing software and hardware framework and makes modest modification of the current system to provide performance improvements comparable to those of the dynamic graph accelerators. 
These accelerators employ graph prefetchers responsible for prefetching neighbors and their property data, which uses a similar strategy as ~\cite{prodigy,metasys,droplet} to prefetch graph data structures and suffer sequential dependency between graph data structures. 
%\amc{} can be integrated into these designs to improve their effective performance further. 

%However, these accelerators often necessitate custom hardware design and programming models, complicating the overall development. Furthermore, due to hardware and software support, it takes additional integration efforts to make it compatible with an existing systems. 

%One major limitation is that dynamic accelerators have not yet been able to run multiple graph applications, resulting in inefficient sharing of hardware resources. The proposed \amc{} leverages the existing software and hardware framework involving modest modification of the current system. \amc{} provides performance improvements comparable to those of the dynamic graph accelerators and simplifies integration and overall development, making it adaptable to a wide variety of applications similar to dynamic graphs.
\section{Motivation and Key Idea}
\label{sec:MOTIVATION}

\begin{figure}[h]
  \centering
  \vspace{-1ex}
  \includegraphics[width=1\linewidth]{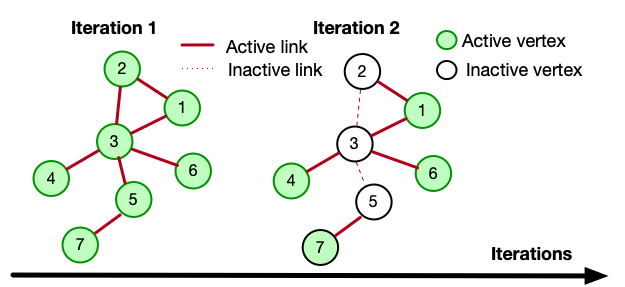}
    % \vspace*{-8mm}
  \caption{Active vertices in PGD across the iterations.} 
  \label{fig:pgd_eg}
  \vspace{-3mm}
\end{figure}

Dynamic graph applications exhibit non-repetitive, irregular memory access patterns. These patterns are difficult to predict using existing prefetchers~\cite{triage,markov,ghb,isb,misb} that use history tables to record and correlate access addresses with either the corresponding PC or previous access addresses. These prefetchers can be categorized as using one-to-many or one-to-one correlations based on the number of accesses linked to a single trigger event. %because of (1) constant change in correlation between PC and access pattern and (2) constant change in adjacent addresses. Therefore, the prefetcher cannot take advantage of PC-address and address-address correlation.
Take PGD~\cite{ligra} as an example. Vertices whose Page Rank value has changed by more than set $\delta$-fraction in previous iteration is active in current iteration. Hence, the set of vertices present in the current iteration will differ from their previous and successor iterations as shown in Fig~\ref{fig:pgd_eg}. In PGD (a push-based algorithm), the vertices send their PageRank value to their neighbors to update their Page Ranks in every iteration. Since the active vertices might change in every iterations, the correlations might also change from its previous iteration. 

%\subsection{Conventional Correlation}
%\label{sec:conventional correlation}

%Conventional prefetchers %depending on the number of accesses linked to a single trigger access.
%Fig.~\ref{fig:pgd_eg} demonstrates an example of changing vertex set in the PGD (a push-based algorithm). 
%These vertices that take part in an iteration are known as active vertices.

%\algrenewcommand\algorithmicrequire{\textbf{Input:}}
%\begin{algorithm}[H]
%\begin{small}
%\caption{Push-based graph traversal}
%\label{alg:push}
%\begin{algorithmic}[1]
%\Require{G(V, E), CurrPR, PrevPR, Frontier}
%    \For{$v \gets 1$ to $V$}
%        \If{$v$ in Frontier} 
%            \For{$n \gets $v to $v+1$}
%            \State {$CurrPR[n]$ $\gets$ {$PrevPR[v]$}}
%            \EndFor
%         \EndIf
%    \EndFor
%\end{algorithmic}
%\end{small}
%\end{algorithm}

For this particular example in Fig~\ref{fig:pgd_eg}, the active vertex set in iteration 1 consists of all the vertices in the graph. The active vertices change to 
%five (1, 2, 4, 5, 7) in iteration 2 and
four (1, 4, 6, 7) in iteration 2. According to the dependency of indirect data structure accesses among the three arrays (V: vertex array, N: neighbor array, P: vertex property array), as shown in Fig~\ref{fig:traversal}, the memory access (misses are marked by *) sequence would look like the following:

\begin{figure}[h]
  \centering
  \vspace{-1ex}
  \includegraphics[width=0.6\linewidth]{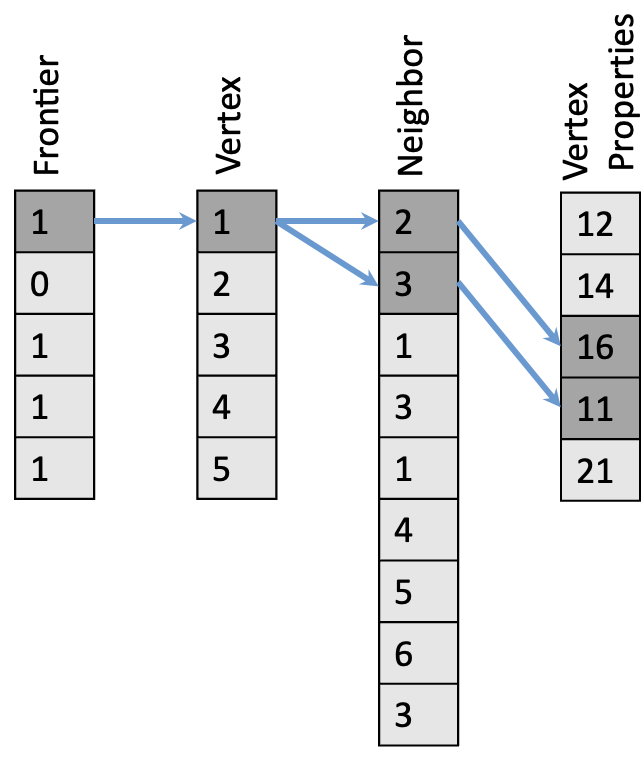}
    % \vspace*{-8mm}
  \caption{{PGD traversal on a graph.}} 
  \label{fig:traversal}
  \vspace{-2mm}
\end{figure}

%For any graph traversal application, the access traversal pattern follows the following high-level indirections between three arrays, namely vertex, neighbor, and vertex property array (vertex->neighbor->vertex property):

\textbf{Iteration 1}: (all vertices are active)
V[1], N[2]*, P[2]*, N[3], P[3]*, V[2], N[1], P[1]*, N[3], P[3]*, V[3], N[4]*, P[4]*, N[5]*, P[5]*, N[6]*, P[6]*, V[4], N[3], P[3]*, ….

% \textbf{Iteration 2}: (vertex 1, 2, 4, 5, 7 are active)
% V[1], N[2], P[2], N[3], P[3], V[2], N[1], P[1], N[3], P[3], V[4], N[3], P[3], …

\textbf{Iteration 2}: (vertex 1, 4, 6, 7 are active)
V[1], N[2]*, P[2]*, N[3]*, P[3]*, V[4], N[3]*, P[3]*, V[6], N[3]*, P[3]*, V[7], N[5]*, P[5]*

The letter b in N[b], which follows V[a], represents the name of vertex a's neighbor. Therefore, vertex b is one of vertex a's neighbors. The address-to-address correlation-based prefetchers~\cite{markov,ghb} records correlation between adjacent addresses during runtime. In iteration 2, on-demand access to vertex 2 to check whether this vertex is active in the current iteration triggers prefetcher, leading to useless prefetching of vertex 2's neighbor (vertex 3) that will not be accessed in iteration 2.
 
\begin{table}[h]{}
\centering
\begin{tabular}{|p{0.5cm}|p{6cm}|}
\hline
 PC & Address Stream \\
 \hline
 A & V[1], V[2], V[3], V[3], V[4], V[5], V[6], V[7] \\
 \hline
 B & N[1], N[2], N[4], N[5], N[6], N[3], N[7], N[5] \\
 \hline
 C & P[1], P[2], P[4], P[5], P[6], P[3], P[7], P[5] \\
 \hline
\end{tabular}
% \vspace{2mm}
\caption{MISB Correlation.}
 \label{tab:MISBCorrelation.}
 \vspace{-4mm}
\end{table}

Recent prefetchers~\cite{triage,misb,isb} combine PC localization with address correlation to build correlations as shown in Table~\ref{tab:MISBCorrelation.}. The accuracy for MISB is 14\%  whereas the coverage is 7\% for covering iteration 2 misses. In this example, it is assumed that each element in the array (V, N, and P) occupies a single cache line for simplicity.

Spatial prefetchers~\cite{vldp,stream,bingo,sms} are limited to record access within the physical page and then prefetch them into the next demand page. Assuming, the vertex, neighbor, vertex property and frontier array lies in four consecutive pages. VLDP develops correlation between the page offsets in OPT and block offset in DPTs (various trigger length) as shown in Table~\ref{tab:VLDPCorrelation.}. Considering Iteration 2 access (shown above) pattern wherein all the accesses to neighbor and vertex property array are L2 misses for baseline with no prefetcher. The accuracy for VLDP is 43\% whereas the coverage is 21\% for covering iteration 2 misses. This shows that MISB suffers from aliasing problem and correlation style of VLDP can overcome this problem to provide comparatively better accuracy. %In order to cover remaining misses accurately, \amc{} develops contextual correlation explained in Section~\ref{sec:contextual correlation}.

\begin{table}[h]
\begin{small}
    \begin{minipage}{.5\linewidth}
      \centering
        \begin{tabular}{|l|l|}
            \hline
            Delta & Prediction \\
            \hline
            1 & 1 \\
            \hline
        \end{tabular}
        \caption*{DPT 1.}
    \end{minipage}%
    \begin{minipage}{.5\linewidth}
      \centering
        \begin{tabular}{|l|l|}
            \hline
            Delta & Prediction \\
            \hline
            1, 1 & 1 \\
            \hline
            1, -2 & 3 \\
            \hline
        \end{tabular}
        \caption*{DPT 2.}
    \end{minipage} 
    \begin{minipage}{.5\linewidth}
      \centering
        \begin{tabular}{|l|l|}
            \hline
            Delta & Prediction \\
            \hline
            1, 1, 1 & 1 \\
            \hline
            1, -2, 3 & 1 \\
            \hline
        \end{tabular}
        \caption*{DPT 3.}
    \end{minipage} 
        \begin{minipage}{.4\linewidth}
      \centering
        \begin{tabular}{|l|l|}
            \hline
            Offset & Prediction \\
            \hline
            1 & 1 \\
            \hline
        \end{tabular}
        \caption*{OPT.}
    \end{minipage}
    \caption{VLDP Correlation.}
    \label{tab:VLDPCorrelation.}
\end{small}
\vspace{-4mm}
\end{table}

%\subsection{\amc{} Contextual Correlation}
%\label{sec:contextual correlation}

With some data structure knowledge, identifying vertex-neighbor correlation is possible during graph traversal. This knowledge can be either provided by application program interface or using compiler analysis~\cite{prodigy}. The \textit{key idea} of \amc{} is to use ``access-to-miss correlation'' between target data structure accesses and other misses to add contextual correlation for correlation-based prefetchers, which can adapt to vertex and edge changes in dynamic graph applications. 
%Existing hardware prefetchers face the problem of aliasing, which can be solved using \textbf{\textit{``many-to-many"}} correlation style history recording. However, VLDP~\cite{vldp} uses \textbf{\textit{``many-to-one"}} correlation style history recording that is limited to a physical page and on-chip hardware table sizes, which leads to poor coverage (Section~\ref{sec:coverage}) as in given example (Fig~\ref{fig:pgd_eg}), in iteration 2 on demand accesses to vertex 1, and 2 prefetcher prefetches vertex 3 which would result in inaccurate prefetching.

These inter-data structure correlations are relatively easy to extract in the source code of dynamic graph applications. Algorithm~\ref{alg:pgd} shows \amc{}\textquotesingle s light-weight interface in PGD application. \amc{} function calls are explained in Table~\ref{tab:amc_func_calls} and Section~\ref{sec:softwareSupport}. First, the programmer needs to identify the target regular data structure, which will act as a trigger agent to record and prefetch stored miss stream. This data structure is mostly a vertex array in graph analytics (delta in PGD). Second, the programmer needs to identify the data structure that accounts for storing active vertices in the iteration (frontier in PGD).

\begin{table}[h]{}
\footnotesize
\centering
\begin{tabular}{|c|c|}
\hline
 Trigger access & Miss stream \\
 \hline
 V[1] & N[2], P[2] , P[3] \\
 \hline
 V[1], V[2] & P[1], P[3] \\
 \hline
 V[2], V[3] & N[4], P[4], P[5], N[6] \\
 \hline
 V[3], V[4] & P[3] \\
 \hline
  V[5], V[6] & N[3] \\
 \hline
  V[6], V[7] & N[5] \\
 \hline
\end{tabular}
% \vspace{2mm}
\caption{\amc{} Correlation Recording.}
\vspace{-4mm}
 \label{tab:amc_record}
\end{table}

\amc{} prefetcher builds \textit{access-to-miss} correlations between L1 target data accesses and L2 misses (excluding L2 target data miss). These L2 misses are the misses that happen in the time frame between two L1 target accesses. L1 target data access is a trigger event to prefetch correlated miss stream associated with it. \amc{} prefetcher observes access patterns in the previous iteration and build correlation entries as shown in Table~\ref{tab:amc_record}. In iteration 2, \amc{} has 60\% accuracy and 43\% coverage over baseline with no prefetcher.
\amc{} prefetcher records virtual addresses of the target data accesses to facilitate faster lookup in \amcCache{} (Section~\ref{sec:amcLookup}) and builds fine-grain correlations between vertex accesses and misses.
Virtual addressing enables \amc{} to lookup \amcCache{} in parallel with the L1 data cache accesses before address translation.
%In addition, \amc{} prefetcher uses the physical address to record the L2 misses, which avoids address translation at the time of prefetching.

% \subsection{~\amc{} Correlation in Real Application\\ Example}
 
\begin{algorithm}[h]
\begin{small}
\caption{PGD using \amc{} prefetcher}
\label{alg:pgd}
\begin{algorithmic}[1]

\Procedure{\textcolor{pink}{Init()}}{}:
    \State Frontier = \{1,...,1\}
    \State Delta = \{$\frac{1}{N}$,...,$\frac{1}{N}$\} 
    %\Comment{\textcolor{codegreen}{Initial delta propagation from each vertex}}
    \State nghSum = \{0, …, 0\}
    \State PR = \{0,...,0\} 
    %\Comment{\textcolor{codegreen}{Stores changes in PageRank values}}
    \State error = $\infty$
    \State \textit{\textcolor{codepurple}{AMC.AddrTBase(Delta, N)}}
    % \Comment{\textcolor{codegreen}{AMC Software interface}}
    \State \textit{\textcolor{codepurple}{AMC.AddrFBase(Frontier, N)}}
    
    \State \textbf{return 1}
\EndProcedure
\State

\Procedure{\textcolor{pink}{Update}}{s, d}: 
%\Comment{\textcolor{codegreen}{Passed to EdgeMap}}
  \State atomic\_increment(nghSum[d], $\frac{Delta[s]}{degree(s)}$)
  \State \textbf{return 1}
\EndProcedure
\State

\Procedure{\textcolor{pink}{Compute}}{i}: 
%\Comment{\textcolor{codegreen}{Passed to VertexMap}}
  \State Delta[i] = $\alpha \times$ nghSum[i]
  \State PR[i] = PR[i] + Delta[i]
  \State \textbf{return (abs(Delta[i] $>$ $\delta$))}
\EndProcedure
\State

\Procedure {\textcolor{pink}{PGD}}{G, $\alpha, \epsilon$}:
  \State \textit{\textcolor{codepurple}{AMC.init()}}
  \State INIT()
  
  \While{(error $>$ $\epsilon$):}
  \State Frontier = \textbf{EDGEMAP}(G, Frontier, \textit{UPDATE})
  \State  Frontier = \textbf{VERTEXMAP}(Frontier, \textit{COMPUTE})
  \State error = sum of nghSum entries;
  \State \textit{\textcolor{codepurple}{AMC.update()}}
  \EndWhile

  \State \textit{\textcolor{codepurple}{AMC.end()}}
  \State \textbf{return PR}
  
\EndProcedure
\end{algorithmic}
\end{small}
\end{algorithm}

% The PGD from ligra~\cite{ligra} is used as an example. As shown in Algorithm~\ref{alg:pgd}, PGD has two data structure types. One is regular (Delta, Frontier, PR), which exhibits a sequential memory access pattern, and another is an irregular (nghSum), which shows an data dependent memory access pattern. In the ``\textit{UPDATE}" procedure, the application access only those vertices in the delta array which are flagged as active vertices in the frontier data structure. This forms an algorithmic \underline{one-to-one} correlation between the frontier and the delta. In PGD, the active vertex set gets updated depending on a condition (Line 18) with every iteration, as shown in Line 25. 
% At Line 12, the application access is only performed to the neighbors (nghSum) of the (active) delta vertex. This forms an algorithmic \underline{one-to-many} correlation between the delta and the nghSum (vertex-neighbors property relationship). Because of changing active vertices in every iteration, the application access to the delta array elements is not repetitive across iterations. However, the access direction is the same (increasing order of indices), and the spatial distance between accesses is significantly less. Therefore, the delta array accesses by application contribute to fewer misses than accesses to the nghSum array, which exhibits irregular memory accesses (data-dependent access).
\section{Software Support for \amc{} prefetcher}
\label{sec:softwareSupport}

This section describes \amc{} functions, architectural state registers, and OS support that are required by the proposed design. An example of using \amc{}\textquotesingle s programming interface for PGD is demonstrated as well.

\begin{table}[h]{}
\begin{small}
% \footnotesize
\centering
\begin{tabular}{|p{2.4cm}|p{5.4cm}|}
\hline
 Function & Definition \\
 \hline
 \amc{}.init() & Set ASID for permission check, allocate memory for \amc{} storage \\
 \hline
\amc{}.AddrFBase (addr, size)  &  Add base address with its corresponding size for frontier data structure \\
 \hline
 \amc{}.AddrTBase (addr, size)  &  Add base address and corresponding size for target data structure  \\
 \hline
 \amc{}.update()  & Set prefetching phase, metadata storage management, resets target access count register\\
 \hline
 \amc{}.end()  & Free \amc{} storage memory space \\
\hline
\end{tabular}
% \vspace{2mm}
\caption{\amc{} Function Calls}
% \vspace{-8mm}
 \label{tab:amc_func_calls}
\vspace{-4mm}
\end{small}
\end{table}

\subsection{\amc{} Functions and Architectural State Registers}
%To establish a contextual correlation between the target accesses in the L1 cache and the L2 cache miss stream, 
\amc{} requires the following additional architectural registers: (1) two pairs of address range registers to hold the start address and size of the target and frontier data structure, (2) a prefetch phase register to enable prefetching after an initial iteration and recycle the off-chip \amc{} space in successive iterations, (3) target access count register, (5) miss count register and (6) four pairs of off-chip \amc{} storage registers to hold the head and tail pointers for the current and the next \amc{} miss addresses and \amc{} index storage.

\amc{} uses the address space identifiers (ASIDs) to distinguish access streams from different processes to do permission checks. The target recorder and frontier buffer in Fig.~\ref{fig:ArchitectureOverview} use a pair of address range registers to filter out the target and frontier accesses from L1 data load accesses. OS allocates off-chip memory space to store both the current and the next \amc{} miss addresses and \amc{} index storage on \textit{\amc{}.init()} function call. \amc{} reserves up to 20\% input size for off-chip \amc{} storage (Section~\ref{sec:offChipStorage}). One can re-purpose unused architectural registers as these special registers for \amc{} prefetcher. An ASID register stores the ASID of the current process using the prefetcher. Target access count register count number of L1 data cache access performed to target data structure (Section~\ref{sec:recordCorrelation}). Miss count register counts the number of L2 misses recorded per \amc{} entry used by compressor unit (Section~\ref{sec:storeCorrelation}). Target access count is used to identify an unique target access during an iteration, whereas the “miss count” counts the number of L2 misses following a target access. 

The \textit{ \amc{}.AddrTBase(addr, size)}, and \textit{ \amc{}.AddrFBase (addr, size)} function provides the system with information of the target and frontier data structure.
The setting of the target address range register happens at the memory allocation time. Using target address range, \amc{} prefetcher can recognize whether access is within the target range (Section~\ref{sec:recordCorrelation}).
\textit{ \amc{}.update()} controls prefetching phase register, metadata storage management, and reset the target access count register (Section~\ref{sec:recordCorrelation}). 
A set state of the prefetching phase register denotes the prefetching is enabled, whereas an unset state of the prefetching is disabled. OS does not disable prefetching except during the context switch (Section~\ref{sec:osSupport}). \textit{\amc{}.end()} function frees up \amc{} off-chip storage and reset all the \amc{} architecture register and invalidate all the entries in \amcCache{} at the end of the execution.

Furthermore, the OS needs to allocate off-chip memory space to store two \amc{} miss addresses and \amc{} index for recording phase and prefetching phase (shown in Fig.~\ref{fig:ArchitectureOverview}). 
One stores the correlation from the previous iteration to perform prefetching and another to learns the correlation in the current iteration.
OS maintains the off-chip address range registers. At the iteration boundary, the \amc{} invalidates its prefetching phase memory space and reuses it to store the upcoming iteration's correlation.  
In short, both recording phase and prefetching phase perform role reversal at the iteration boundary.

\subsection{OS Extension}
\label{sec:osSupport}
The OS is responsible for the process management, interrupts service, I/O, virtualization, and resource management of different cores in the system. In case of long latency events such as page faults or interrupt service routine, the OS needs to switch out the current process, known as context switch, and handle the event. Conventional prefetchers either flush the metadata entries or save them in memory to retrieve them later. \amc{} prefetcher can reuse its old metadata after being context-switched back again only when there is no swapping of physical pages from the process. During a context switch, the physical pages swap when memory runs out of physical pages to allocate to the new context switch process, which is not typical. Suppose the page's swap, the \amc{} resets its metadata, disables prefetching, and restarts from the recording phase. The dynamic graph applications consist of multiple iterations. Therefore, \amc{} can quickly perform its recording phase in the current iteration and start prefetching in the next iteration.

\subsection{An Example of using \amc{}\textquotesingle s Programming Interface}
The PGD algorithm from Ligra~\cite{ligra} suite is modified to demonstrate how to use \amc{}\textquotesingle s programming interface (Algorithm~\ref{alg:pgd}). 
Line 1 - 9 initializes the data structures used in the algorithm. 
Line 11 - 13 calculates the page rank value for the vertex present in frontier (active vertex in current iteration). Line 15 - 18 calculates the set of vertices for the next iteration.
Purple-colored function calls are \amc{} functions. Line 21 initializes the \amc{} prefetcher registers, allocates off-chip memory for \amc{} metadata storage, and resets all of the \amcCache{} entries (Section~\ref{sec:architectureDescription}) as well as all of the architectural state registers. 
Line 7 and 8 defines the virtual address range for the target data structure and the frontier with their corresponding size (N is the number of elements). 
%Line 8  defines the virtual address range for the frontier with its size.

Line 27 denotes the boundary of an iteration. It sets the prefetching phase register to enable prefetching after the initial iteration. Additionally, after every iteration, it invalidates the off-chip \amc{} storage used for prefetching miss stream for the current iteration and reuses it to record in the next iteration. It does not invalidate the correlations recorded in the current iteration. Finally, line 28 terminates the \amc{} prefetcher and frees up its off-chip memory space.

\section{\amc{} architecture}
\label{sec:architectureDescription}

\begin{figure*}
\centering
\includegraphics[width=\linewidth,height=10cm,keepaspectratio]{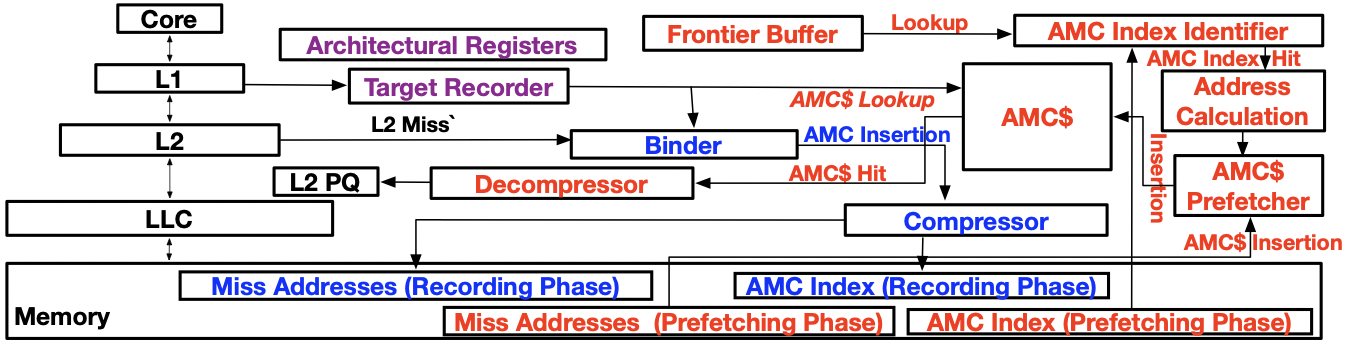}
  % \vspace*{-8mm}
\caption{An overview of proposed \amc{} architecture. These blocks are private to the core.}. 
\label{fig:ArchitectureOverview}
\vspace{-8mm}
\end{figure*}

The \amc{} prefetcher adds a few architectural components to a conventional cache hierarchy as shown in Fig~\ref{fig:ArchitectureOverview}.
%Fig.~\ref{fig:ArchitectureOverview} shows an overview of proposed \amc{} architecture. 
Binder, Compressor, and off-chip storage space are used during correlation recording explained in Section~\ref{sec:recordCorrelation} and \ref{sec:storeCorrelation}. Frontier buffer, Target Recorder, \amc{} Index Identifier, \amcCache{} Prefetcher, \amcCache{}, and Decompressor are the components used during prefetching explained in Section~\ref{sec:amcCache}. Architectural Registers and Target Recorder are the common components between building correlation and prefetching.
 %Each of the following subsections discusses a group of components for recording correlations, storing correlations in memory, and caching correlations on-chip. %To take advantage of the dependency between data structures and efficiently architect in terms of on-chip area, power consumption, and performance. The proposed work lays its design decisions on the following questions: 

\subsection{Correlation Recording}
\label{sec:recordCorrelation}
\amc{} records many-to-many correlations between accesses to a target data structure and L2 misses. Target data accesses identified in the L1 cache act as trigger events to prefetch to the L2 cache. Hence, the \amc{} needs to record target data structure's accesses at the L1 data cache and the following misses from the L2 cache to build the correlations. An \amc{} correlation entry consists of two target accesses (2$\times$64 bits) and up to 20 misses (20$\times$46 bits). A \textit{Target Recorder} is used to identify the target L1 accesses, which can hold up to two most recent target accesses. The target recorder includes a target access counter that increments on every demand target access to the L1 data cache. This counter is reset when the AMC.update() function is invoked.  %`Target Recorder' uses a FIFO replacement scheme.

The L2 misses that do not belong to the target address range are tagged with the latest target access count value at the time of the L2 miss and forwarded to a \textit{Binder} to build a correlation entry. A \textit{Miss Count} holds the number of misses belonging to the same target access count. When a miss with a different target access count value arrives at the Binder or when the Miss Count reaches 20, this entry is compressed and sent to memory, the Miss Count is reset, and a new correlation entry is initiated. The access count retrieves the correlated target accesses in the Target Recorder.

%Then the \amc{} access the Target Recorder to extract the entry belonging to the completed recording using the corresponding target access count. Finally, the matched entry institute correlation to the recorded miss stream. If there are sequential miss streams of more than 20 misses belonging to the same target recorder, \amc{} prefetcher stores the following miss streams in a group of 20 together in an \amc{} entry. \amc{} records the latest target access from the extracted target recorder entry as one of the target access for the next target recorder entry. Therefore, except the very first \amc{} entry in an application, all the \amc{} entries consist of two target accesses. Section~\ref{sec:missSizeStdy} performs a sensitivity study on miss size to suggest why 20 is an appropriate miss stream length choice. 
%\amc{} prefetcher writebacks compressed form of this entry to off-chip recording phase metadata (\amc{} miss addresses and \amc{} Index).

\subsection{Storing Correlations}
\label{sec:storeCorrelation}

\begin{figure}
  \centering
  \vspace{-1ex}
\includegraphics[width=0.8\linewidth]{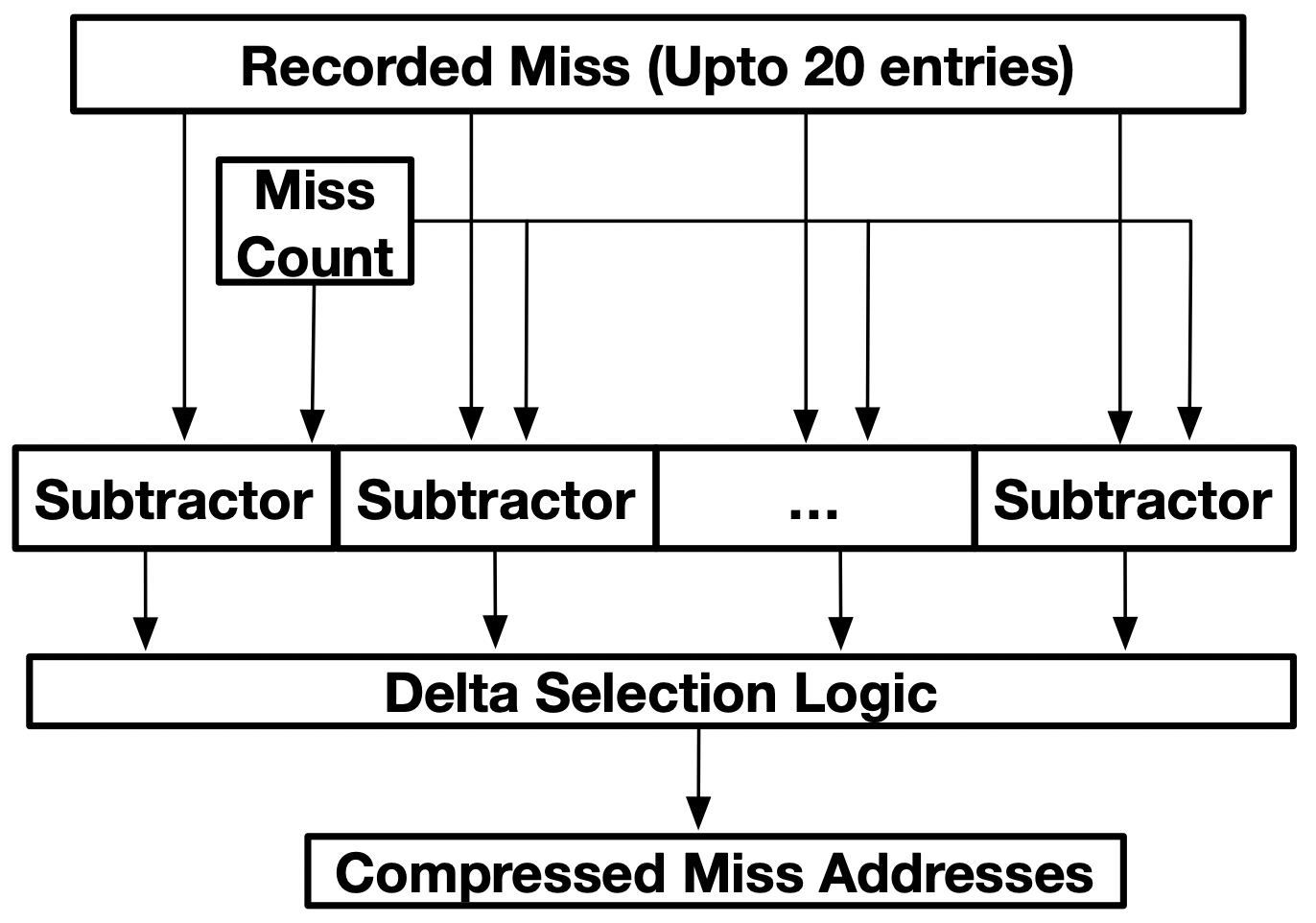}
\caption{ Compressor Design.}
\label{fig:compressor}
\vspace{-5mm}
\end{figure}

The \amc{} prefetcher maintains two off-chip metadata storage simultaneously: one for recording phase, the other for prefetching. Each of these two metadata storage contains two tables: Miss Addresses and \amc{} Index. \amc{} prefetcher stores the correlations learned during the current iteration and uses the correlations learned from the last iteration for prefetching. At the end of every iteration when \textit{AMC.update()} is evoked, the head pointers of these two memory spaces are swapped to allow the latest correlations to be used for prefetching while recycling the memory space for recording new correlations. %phase metadata storage to perform prefetching in the current iteration. 
This continuous learning while prefetching helps to capture the changes in vertices and edges of the dynamic graphs. % during runtime. This adaption is at the iteration level, and the new correlation formed in the current iteration applies in the next successive iteration.

The compressed miss addresses store miss address entries with different sizes in compact format. When a new entry is compressed and sent to memory. The tail pointer of Miss Addresses of recording phase is incremented based on this entry's compressed size for storing the next entry.
The \amc{} Index store the target addresses and metadata of each correlation entry. Each \amc{} Index entry consists of two target addresses, compression mode, a pointer to the Miss Addresses, and the number of correlated misses. 
%Pragmatically, 100 entries are enough to match the pace of dynamic change for the selected applications. 

\begin{figure}
  \centering
  \vspace{-1ex}
  \includegraphics[width=1\linewidth]{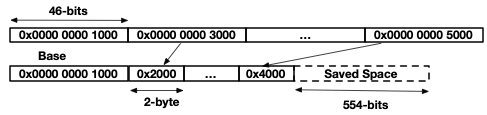}
  \caption{ 46-bit base 2-byte-$\Delta$ compression example.}
\label{fig:compressorEx2byte}
\vspace*{-2mm}
\end{figure}

A lightweight Base$\Delta$ compression variant~\cite{baseDeltaCom} is designed to save off-chip bandwidth and reduce metadata storage.  An \amc{} entry stores up to 20 physical addresses (52-bit physical memory) without block offset  of the misses (20$\times$46 bit without compression). \amc{} uses the physical block address of the first miss as the base (46 bit) and three different sizes of deltas (1, 2, or 4 byte) to compress the other misses. A 2-byte-$\Delta$ example is shown in Fig~\ref{fig:compressorEx2byte}. When all of the addresses can be represented with Base+$\Delta$, this entry can be compressed with the corresponding size of delta. 
%, fig:compressorEx4byte, fig:compressorEx2byte}.
Fig~\ref{fig:compressor} illustrates a high-level view of the compressor design.
The compressor uses the Miss Count to activate the number subtraction units equal to the number of misses. Three delta sizes are tested in parallel. The smallest compressable delta size is selected using delta selection logic. 
%The functionality of \textit{Compressor Selection} \textit{compressor or not} (CoN) is the same as B$\Delta$I~\cite{baseDeltaCom}. It selects the compressed metadata which occupies the smallest size.

%After compression, the following metadata fields are added to each recorded entry. The compressor unit uses the Miss Count to set the subtraction amount. A \textit{Compressed-or-Not Flag} indicates whether the stored miss stream is compressed. A \textit{Compressor Select} field selects the compression that occupies the smallest size. 
% The above explaination is confusing, I don't understand how does the compression work.

%\begin{figure}[h]
%  \centering
%  \vspace{-1ex}
%  \includegraphics[width=1\linewidth]{Images/baseDeltaExam.png}
%  \caption{ 46-bit base 1-byte-$\Delta$ compression example.}
%\label{fig:compressorEx}
%\end{figure}

%\begin{figure}[h]
%  \centering
%  \vspace{-1ex}
%  \includegraphics[width=1\linewidth]{Images/baseDeltaExam4Byte.png}
%  \caption{ 46-bit base 4-byte-$\Delta$ compression example.}
%\label{fig:compressorEx4byte}
%\end{figure}

The target access addresses do not undergo compression because the target addresses play a critical role in \amcCache{}, which will be explained in Section~\ref{sec:amcLookup} and Section~\ref{sec:metaPrefetching}.
Compressing the target access addresses adds delay to the critical path and might lead to late prefetches. %Therefore, adding a separate compressor/decompressor unit does not justify the benefit of saving on-chip and off-chip storage space. 
Instead, only the delta of the target accesses is recorded by the target recorder using target start address stored in the architectural address range register (Section~\ref{sec:softwareSupport}).
For the evaluated workloads, the compression ratios for 20 recorded uncompressed misses using different deltas are 4.5 (920/206) for 1-byte-$\Delta$ (best-case), 2.51 (920/366) for 2-byte-$\Delta$, and 1.34 (920/686) for 4-byte-$\Delta$ (worst-case). 
%If an uncompressed correlated \amc{} entry consists of multiple 20 miss streams, each miss stream performs individual compression.

\subsection{\amcCache{}}
\label{sec:amcCache}

\begin{figure}[h]
  \centering
  \includegraphics[width=1\linewidth]{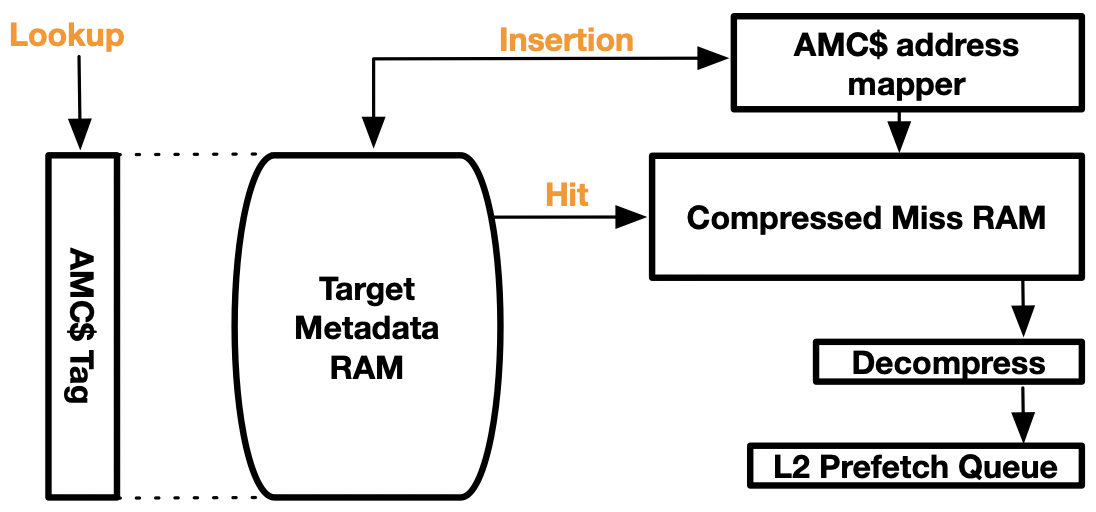}
    % \vspace*{-8mm}
  \caption{An illustration of \amcCache{}.} 
  \label{fig:amcCache}
   \vspace*{-2mm}
\end{figure}

\amc{} uses an on-chip cache to prefetch the access-to-miss address correlations recorded during the last iteration. These correlation entries are inserted into the cache in sequence and evicted in FIFO order. 
Fig~\ref{fig:amcCache} illustrates a high-level view of the \amcCache{} design. The three main components of \amcCache{} are (1) \amcCache{}'s tag, storing target addresses  as the tags for cache lookup, (2) target metadata RAM, storing the location of the corresponding entry in the \amcCache{}, and (3) a Compressed Miss RAM, stores the compressed misses.

%\amc{} allows the same target address stream to store multiple 20 miss streams in compressed miss RAM. Choosing the \textit{RAM} structure over associative enables the miss streams of the same \amcCache{} tag entry to be contiguous and extract misses in a streaming fashion. It reduces the overall hardware overhead of the associative replacement mechanism.

%The three main mechanisms of on-chip \amcCache{} are:

\subsubsection{\amcCache{} Lookup}
\label{sec:amcLookup}
\amcCache{} tag is a content-addressable memory (CAM) \cite{cam} that allows comparing a target address. In this case, \amc{} uses the addresses identified by the Target Recorder to lookup the \amcCache{} tag. A matched tag returns the associated Target Metadata RAM entry.
Each entry in the Target Metadata RAM consists of a valid bit (1 bit), the number of misses in the entry (5 bits), the compression mode (2 bits: 1/2/4-byte-$\Delta$), and the head pointer of miss addresses stored in the Compressed Miss RAM. 
%When an entry in Target recorder is assigned the target access count value, it is ready to look up \amcCache{}. \amc{} uses the latest access in the target access stream to look up \amcCache{} tag. 
On an \amcCache{} hit, \amc{} uses the number of misses and the head pointer to extract the corresponding compressed miss addresses stored in the Compressed Miss RAM. %; if no entry matches, it's a \amcCache{} miss.
\amc{} prefetcher passes these compressed miss addresses to a Base$\Delta$ decompressor to generate L2 prefetch candidates. 
It is possible to have multiple hits in the \amcCache{} tag because a correlation with more than 20 misses can be split into multiple correlation entries.  In case of multiple hits, \amc{} extracts corresponding entries one-by-one to decompress the miss addresses. 
\amcCache{} hit entries are written back to off-chip recording phase metadata storage for the next iteration.

\subsubsection{\amcCache{} Insertion}
\label{sec:metaPrefetching}
\amc{} keeps track of the processing progress of the frontier array to issue timely prefetches. 
A \textit{Frontier Buffer} is used to identify and record the addresses of the frontier accesses, similar to the Target Recorder. These frontier addresses are then used to determine when to prefetch \amc{} Index entries to the \textit{\amc{} Index Identifier}. The \amc{} Index Identifier caches a continuous subset of the \amc{} Index entries from the off-chip \amc{} Index.
When two frontier accesses record to an entry of frontier buffer, \amc{} uses frontier deltas to lookup \amc{} Index Identifier. Frontier deltas are obtained by subtracting frontier access address with frontier start address stored in architectural address range register. The \textit{Address Calculation} aligns the frontier delta with the target delta size to obtain the corresponding target delta using this equation target\_delta = frontier\_delta $\times$ (target\_size/frontier\_size). The frontier and target size are obtained using architectural registers.
%The frontier buffer tries to capture the active vertices that will traverse in the current iteration.
%\amc{} Index Identifier holds multiple entries consisting of target access stream from off-chip \amc{} Index (prefetching phase metadata). Since the frontier and target data structure access the same active vertex in the iteration, the delta offset of the frontier and target data array is precisely the same.

On an \amc{} Index Identifier hit, \amcCache{} prefetcher prefetches the corresponding entry from off-chip \amc{} Miss Addresses to the \amcCache{}. 
\amc{} uses the pointer to the entry in the Miss Addresses, compression mode, and the number of misses to prefetch the miss addresses.
The Target Metadata RAM stores the hit entry of the \amc{} Index Identifier with a pointer to the Compressed Miss RAM that is copied from the tail pointer of the Compressed Miss RAM.
The miss addresses are stored at the current tail pointer of Compressed Miss RAM, which is then advanced based on size of the compressed misses for the next entry.
The \amc{} Index Identifier invalidates all previous entries, including the hit entry, to fill up the \amc{} Index Identifier with the next set of \amc{} Index entries. %The location of the hit entry within the range decides whether the prefetched metadata will lead to timely prefetching. For example, the hit entry closer to the start range \amc{} Index Identifier means an early prefetch, whereas a hit near the end range means a late prefetch.

The purpose of \amc{} Index Identifier is to prefetch metadata on-chip to reduce the delay of prefetching \amc{} entries to \amcCache{}. The \amc{} Index Identifier stores a range of Target entries from off-chip \amc{} Index. Once the set range of \amc{} Index entries is populated to \amc{} Index Identifier, the head pointer points to the past last entry being fetched into \amc{} Index Identifier for the next refill.
On an \amc{} Index Identifier miss, if the latest frontier delta is greater than the latest delta of the last entry of the \amc{} Index Identifier, the entries of the \amc{} Index Identifier are invalidated and replaced by next batch of target entries from off-chip \amc{} Index. %using the head pointer of the \amc{} Index.

\subsubsection{\amcCache{} Replacement}

\amcCache{} uses a FIFO replacement to invalidate earlier entries when there are no invalid entries in the \amcCache{} tag or no space in the Compressed Miss RAM. This simplifies the cache design. There is no eviction from \amcCache{} to the off-chip metadata storage.
\section{Experimental setup}
\label{sec:setup}

\begin{table}[h]{}
\begin{small}
%\centering
\begin{tabular}{|p{1.6cm}|p{6cm}|}
\hline
\textbf{Core Parameters} & 4 OoO cores, 4GHz, 4-wide, 256 ROB, 64 LQ, 64 SQ, perceptron branch predictor~\cite{branch} \\

\hline

\textbf{L1 D/ICache} & private, 64KB, 8-way, 4 cycles, 64B block, MSHR: 8\\ 
\hline

\textbf{L2 Cache} & private, 256KB, 8-way, 12 cycles, 64B block, MSHR: 16,  next-line prefetcher\\

\hline
\textbf{LLC Cache} & shared, 8MB, 16-way, 42 cycles, 64B block, MSHR: 128 \\
\hline

\textbf{Memory Controller} & FCFS, read queue size = 64, write queue size = 32 write queue draining: high/low threshold = 75\%/25\%\\
\hline
\textbf{Main Memory} & DDR4, 8Gb (x16 I/Os), 2400 MT/s, 1 channel, 1 rank, 16 banks, tRCD = tRP = tCL = 17 cycles\\
\hline

\end{tabular}
% \vspace{-8mm}
\caption{Processor configuration (baseline)}
\vspace{-4mm}
\label{tab:amc_base_conf}
\end{small}
\end{table}

This work uses ChampSim~\cite{champsim}, a trace-based simulation infrastructure, to evaluate the performance of the proposed \amc{}. ChampSim has been used in prefetching competitions. ChampSim's cache system implements FIFO read and prefetch queues. It accurately models bank and bus contention, page table, TLB caches, and TLB functions such as page table walks. The core parameters are modeled based on Intel i7-6700 \cite{corepara} and shown in Table~\ref{tab:amc_base_conf}. The memory timing constraint comes from Micron MT40A2G4 DDR4-2400-CL17 data sheet~\cite{dramcfg}.
The on-chip area and energy consumption of Base$\Delta$ Compressor are estimated using 45nm Synopsys standard cell library~\cite{synopsys} (RTL synthesis) and scaled down to 22nm. To realize the energy benefits of the proposed work, we developed an analytical model based on McPAT~\cite{mcpat}, CACTI~\cite{cacti}, and Micron DDR4 SDRAM SystemPower calculator~\cite{ddr4power}. CACTI is used to get per-access energy for different levels of cache. McPAT is used to get the energy consumed by the core. We modify the Micron DDR4 SDRAM System-Power calculator to model memory energy consumption with current numbers from Micron MT40A2G4. ChampSim does not model OS implications of context switches.

This work evaluated common dynamic graph kernels~\cite{ligra} and real-word graphs~\cite{datasets} that are run until completion. The dynamic graph kernel PGD and Connected Components (CC) are from Ligra. This work modifies BFS and BellmanFord kernels from Ligra~\cite{ligra} using a strategy similar to \cite{tdgraph,graphbolt} i.e., these kernels are simulated twice with two different inputs to create a dynamic graph situation similar to existing techniques~\cite{tdgraph,graphbolt,tornado,vaziri2021controlling,vora2017kickstarter}. For the first time, 80\% of the vertices are randomly selected; for the second time, 10\% of vertices from the first input graph are randomly deleted and 10\% of vertices from the original input are added.

\begin{table}[h]{}
\centering
\begin{small}
\begin{tabular}{|p{1.4cm}|p{1.2cm}|p{1.2cm}|p{0.8cm}|p{2cm}|}
\hline
Datasets & Vertex (Million)  & Edges (Million) & Degree & Type \\
\hline
Amazon  & 0.4 & 3.39 & 9 &  Product network \\
\hline
Stanford  & 0.28 & 2.31 & 9 &  Web graph Stanford \\
\hline
Youtube  & 1.16 & 2.99 & 3 &  Online social network \\
\hline
Road-CA  & 1.97 & 5.53 & 3 & Road network California \\
\hline
ComDblp & 0.43 & 0.36 & 1 & DBLP collaboration network \\
\hline
Google  & 0.88 & 5.11 & 6 & Web graph Google \\
\hline
NotreDame & 0.33 & 1.5 & 5 & Web graph Notre Dame \\
\hline
\end{tabular}
\end{small}
% \vspace{-8mm}
\caption{Input Datasets~\cite{datasets}, }
\vspace{-4mm}
\label{tab:input_info}
\end{table}

% \begin{table}[h]{}
% \centering
% \begin{small}
% \begin{tabular}{|p{1.2cm}|p{1.2cm}|p{1.2cm}|p{0.8cm}|p{0.8cm}|p{1.2cm}|}
% \hline
% Datasets & Vertex (Million)  & Edges (Million) & Degree & Size (MB*) & Type \\
% \hline
% Amazon  & 0.4 & 3.39 & 9 & 24 & Product network \\
% \hline
% Stanford  & 0.28 & 2.31 & 9 & 17 & Web graph Stanford \\
% \hline
% Youtube  & 1.16 & 2.99 & 3 & 29 & Online social network \\
% \hline
% Road-CA  & 1.97 & 5.53 & 3 & 54 & Road network California \\
% \hline
% ComDblp & 0.43 & 0.36 & 1 & 5.1 & DBLP collaboration network \\
% \hline
% Google  & 0.88 & 5.11 & 6 & 41 & Web graph Google \\
% \hline
% NotreDame & 0.33 & 1.5 & 5 & 1.5 & Web graph Notre Dame \\
% \hline
% \end{tabular}
% \end{small}
% % \vspace{-8mm}
% \caption{Input Datasets~\cite{datasets}, }
% % \vspace{-2ex}
% \label{tab:input_info}
% \end{table}

Table~\ref{tab:input_info} lists the real-world data sets used for evaluation. The simulation setup uses different data sets for all the kernels. This different input set is because a few inputs, e.g., Road-CA for PGD, require weeks to finish.  All the evaluated kernels use the Single Program Multiple Data model~\cite{spmd} similar to RnR~\cite{rnr}. Every task executes the same program along with graph partitioning~\cite{partition, shun2016parallel}. In the evaluated simulation setup, the master process is responsible for initializing all the data structures in the algorithm and partitioning the graph into four partitions using METIS~\cite{partition}. These partitions are assigned to each worker to process to perform their computation. Once the worker completes all the computations, the master process collects the updated data structures and finishes the overall graph analysis. 

\begin{table}[h]{}
\begin{small}
\centering
\begin{tabular}{|p{1.4cm}|p{0.8cm}|p{5.4cm}|}
\hline
\textbf{Bingo~\cite{bingo}} & 119kB  & 16K entry history table, degree: 32 \\
\hline
\textbf{VLDP~\cite{vldp}} & 998B & OPT 128B, DHB 222B, DPT 648B, degree: 4 \\
\hline
\textbf{RnR~\cite{rnr}} & 1KB & Window size 512, Buffer size = 256, degree: 512  \\
\hline
\textbf{MISB~\cite{misb}} & 49kB  & 32kB cache, 17kB bloom filter, degree: 32 \\
\hline
\textbf{AMC24kB} & 29kB & 24kB \amcCache{}, 5kB Base$\Delta$ compressor, 100-entry target recorder, 100-entry \amc{} index identifier, 100-entry frontier buffer, degree: correlated stream \\
\hline
\end{tabular}
% \vspace{-8mm}
\caption{On-Chip Storage cost of evaluated prefetchers.}
  \vspace{-4mm}
\label{tab:prefetcherConfig}
\end{small}
\end{table}
\section{Evaluation results}

\begin{figure*}[h]
\centering
\includegraphics[width=\linewidth]{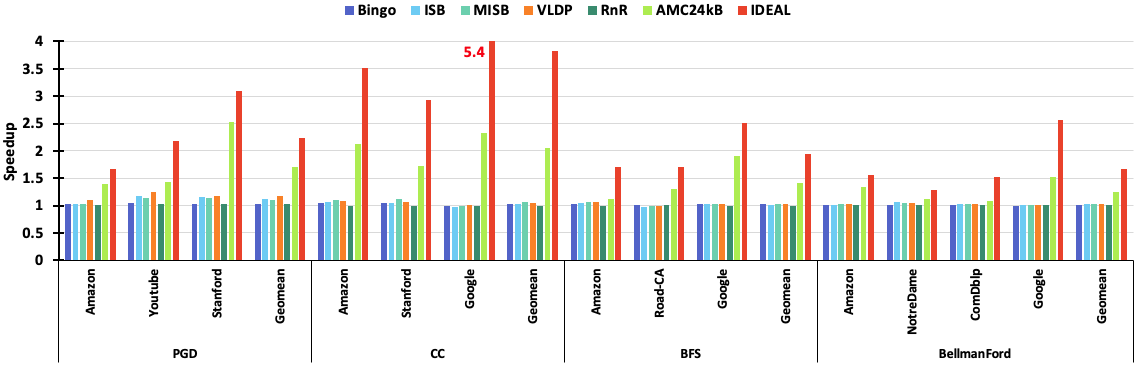}
  % \vspace*{-8mm}
\caption{Speedup over baseline configuration (Table~\ref{tab:amc_base_conf}).}
\label{fig:speedup}
\vspace{-4mm}
\end{figure*}
\label{sec:evaluation}

The baseline system uses the next-line prefetcher as the L2 data prefetcher to evaluate \amc{} prefetcher and other prefetchers. The modern systems ~\cite{composite_pref} employs composite prefetcher \cite{dol,bouquet} to target different access patterns in the kernel. Therefore, it becomes important to evaluate prefetcher design in composite settings. No specific data structure range is assigned to the next line prefetcher to keep the environment as close to reality as possible. \amc{} prefetcher is compared against five prefetchers: (1) Bingo~\cite{bingo}, (2) ISB ~\cite{isb}, (3) MISB~\cite{misb}, (4) VLDP~\cite{vldp}, and (5) RnR~\cite{rnr}. All these prefetchers are trained on L1 data cache access/miss and assigned as L2 prefetcher except RnR. RnR prefetcher is trained at L2 as presented in its original proposal to have a fair comparison with similar software-assisted hardware prefetcher.

Table~\ref{tab:prefetcherConfig} describes the configurations of all these prefetchers. With the exception, the configuration of ISB~\cite{isb} uses an ideal on-chip metadata cache with zero access latency, and infinite size, the degree of the prefetching set to the number of correlated stream lengths. The ``IDEAL'' case is analyzed by having an infinite-sized L2 cache.
As \amc{} prefetcher uses more than one trigger access to initiate prefetching, this work is compared to VLDP and BINGO, which uses a similar lookup strategy to solve the aliasing problem with temporal prefetchers.
Ultimately, we compare it to RnR, which lies in the same category that achieves close to 100\% accuracy on long repeating irregular kernels.

\subsection{Performance}
\subsubsection{Speedup}

The speedup is defined as $\frac{PrefetcherIPC}{BaselineIPC}$ (Fig~\ref{fig:speedup}). The proposed work evaluates BFS and BellmanFord on the second run when the input graph changes similar to previous dynamic graph accelerators~\cite{tdgraph,graphbolt,tornado,vaziri2021controlling,vora2017kickstarter}. The total number of iterations for PGD and CC depends on graph input. The initial iteration traverses all the vertices in the graph for PGD and CC. The vertices become active for the next iteration depending on the vertex property value. 
\amc{}`s off-chip metadata storage recycles metadata that will not be used in future iterations and breaks the dependency between application footprint and metadata size. Additionally, \amc{} uses multiple trigger access similar to BINGO and VLDP to form accurate correlations and differentiate between similar patterns that ISB/MISB cannot. For PGD, and CC \amc{} prefetcher performs 1.71$\times$, 2.04$\times$ (geomean) respectively better than baseline whereas VLDP performs 1.17$\times$ and 1.05$\times$ respectively.

\amc{} prefetcher becomes more accurate in correlating miss streams with iterations because the concurrent recording phase creates a new correlation with every iteration. 
\amc{} does not rely on histories beyond the last iteration. This is because evolving graphs do not typically change rapidly, and the two adjacent iterations have enough similarities. \amc{} prefetches all the addresses except the target data structures because they are vertex arrays that are contiguous, hence the address range is bonded. These data structures can be prefetched using the next-line prefetcher, which is separated from the \amc{} prefetcher. 

For BFS and BellmanFord, the performance improvement could be better than the PGD and CC. This low performance improvement is because the end-to-end evaluation consists of only two instances. With more instances, performance improvement will increase.
\amc{} prefetcher performs about 1.40$\times$ and 1.25$\times$ (geomean) better than baseline whereas VLDP performs 1.14$\times$ and 1.10$\times$ for BFS and BellmanFord respectively. 
As RnR works for long repetitive iterations, the dynamic graph will only behave close to static when the percentage change of vertex/edge added/subtracted is marginally small. RnR performs marginally better than the baseline. The primary reason \amc{} is better than RnR~\cite{rnr} in handling dynamic graphs is its adaptability. Unlike RnR, which replays the same recorded irregular memory access pattern from the initial iteration, \amc{} updates its association table for every iteration.

Through a quantitative comparison, the RnR~\cite{rnr} paper analyzes DROPLET's~\cite{droplet} prefetching strategy on timeliness. DROPLET and PRODIGY~\cite{prodigy} require access to the value to calculate the next prefetching candidate's address. This dependency causes prefetching delay. We use the DROPLET model similar to the RnR paper to model PRODIGY, and AMC performs about 1.56X (geomean) better than PRODIGY.
Prodigy has pointed out that it cannot account for additional control-flow information that leads to cache thrashing~\cite{prodigy}.
Domino's~\cite{bakhshalipour2018domino} (many-to-many correlation) performance is worse than MISB. Quantitatively AMC performs 1.6x (geomean) better than Domino (degree: 4). Since Prodigy, Domino performs worse than the baseline and is therefore excluded from further evaluation.

\subsubsection{Miss Coverage}
\label{sec:coverage}
\begin{figure}[h]
  \centering
  \includegraphics[width=\columnwidth]{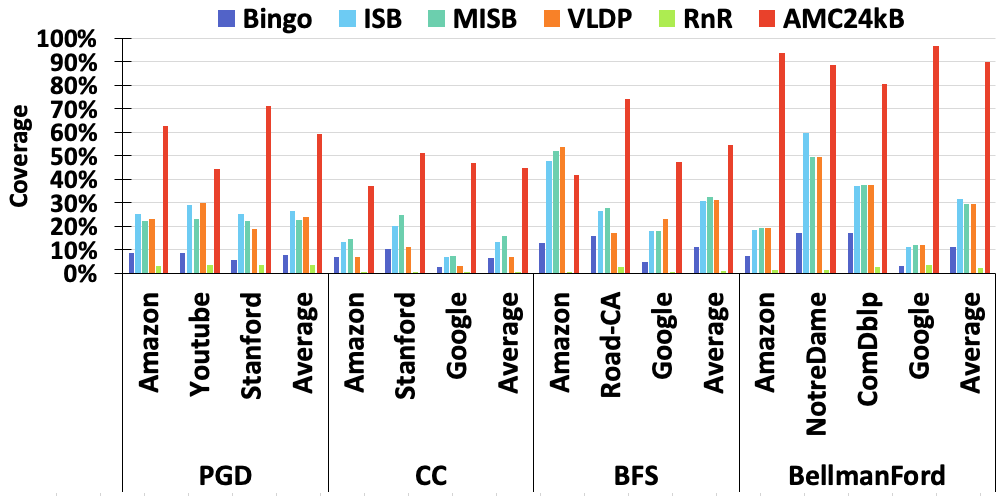}
  % \vspace*{-8mm}
  \caption{L2 miss coverage.} 
  \label{fig:coverage}
\vspace{-4mm}
\end{figure}

The coverage is $\frac{UsefulPrefetchers}{TotalBaselineMisses}$ (Fig~\ref{fig:coverage}). It refers to the total number of baseline misses covered by the prefetcher. It conveys the prefetcher's effectiveness in predicting upcoming misses.
The \amc{} prefetcher's uncovered miss in the iteration is due to changing active vertex set and cold miss in the initial iteration. \amc{} prefetcher on average covers about 59.43\% and 45\% misses in L2. 
For BFS and BellmanFord, the coverage improvement is much more significant than PGD and CC. 

The active vertex set for graph traversal for Bellmanford change marginally (2-7\%); therefore, \amc{} prefetcher acts as RnR on static graph. \amc{} prefetcher covers on an average about 54.51\% and 89.95\% L2 misses. The exception here is amazon input for BFS, which covers about 10\% less than MISB. This exception is because in BFS, if the parent node gets changed, the whole graph traversal changes, and thus, the recorded miss stream by \amc{} becomes useless. Overall, \amc{} performs better than MISB on this input for BFS because even if the recorded miss stream will not be the demand access stream in the next stream. \amc{} do not issue those miss streams and therefore have high accuracy (Section~\ref{sec:accuracy}) and do not cause cache pollution. Spatial prefetchers (VLDP and Bingo) have low coverage because the graph does not exhibit spatial location due to their large size and data-dependent accesses. 
RnR suffers from low coverage (1.7\%) because successive iterations do not have the exact same memory access pattern correlated with irregular access count as the initial iteration.

\subsubsection{Accuracy}
\label{sec:accuracy}

\begin{figure}[h]
  \centering
  % \vspace{-1ex}  
  \includegraphics[width=1\columnwidth]{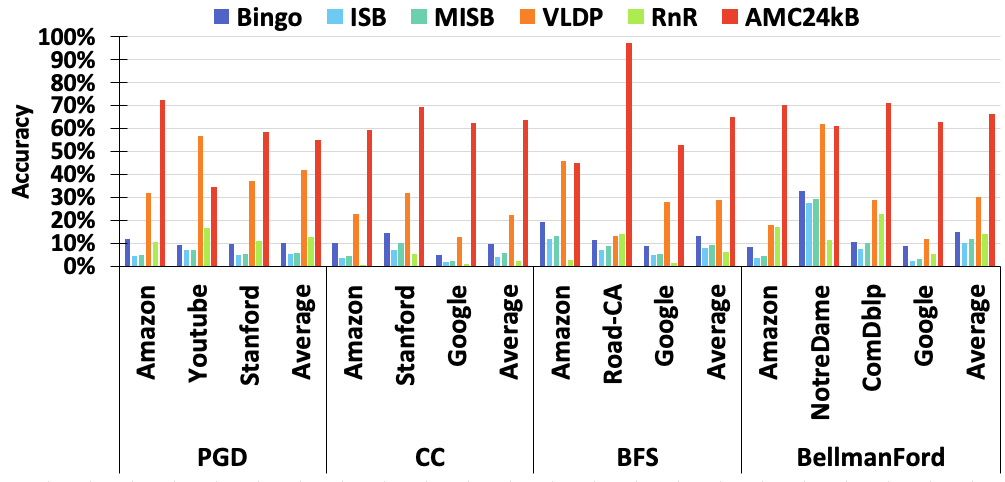}
  % \vspace*{-8mm}
  \caption{Prefetch accuracy.}
  \label{fig:accuracy}
  \vspace{-4mm}
\end{figure}

The accuracy is $\frac{UsefulPrefetchers}{TotalPrefetchers}$ (Fig~\ref{fig:accuracy}). It is the ratio of useful prefetchers to the total number of prefetchers issued. The useful prefetchers bring the cache block into the cache level before its demand access arrives. As \amc{} prefetcher uses access to miss correlation along with the many-to-many correlation style, the probability of issuing inaccurate prefetchers reduces. On average, \amc{} prefetcher achieves an accuracy of 55\%, 63.7\%, 65\%, and 66.4\% for PGD, CC, BFS, and BellmanFord, respectively.

Similarly, VLDP uses many-to-one correlation to issue accurate prefetchers and, compared to other prefetchers, has better accuracy of 31\%. In the case of PGD with Youtube input, the accuracy of VLDP is much better than \amc{}. This indicates that having the finer granularity of correlation (within a page) works better for some graph layouts than vertex-vertex dependent correlations.

\subsubsection{Timeliness}
\label{sec:timeliness}

\begin{figure}[h]
  \centering
  \includegraphics[width=1\columnwidth]{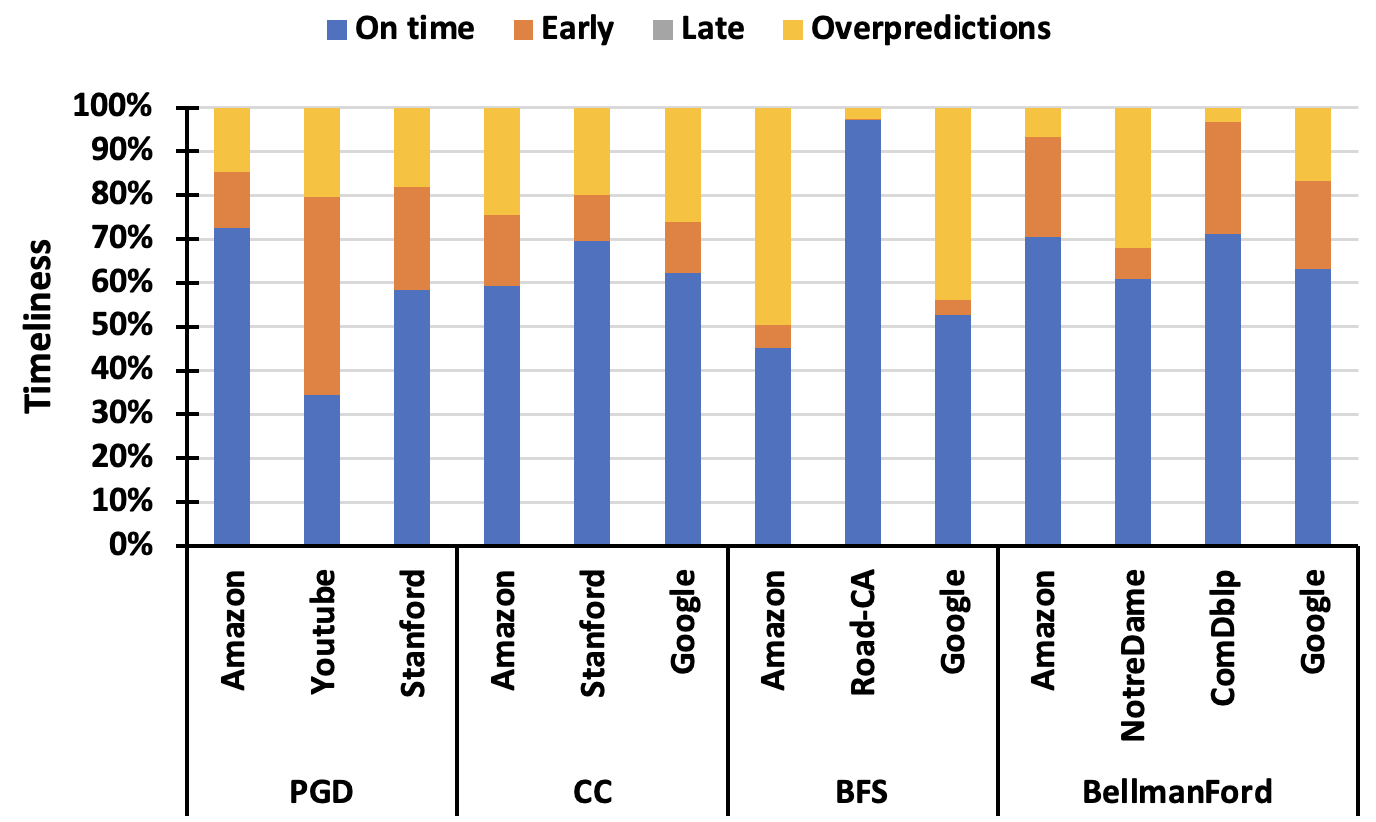}
  % \vspace*{-8mm}
  \caption{\amc{} timeliness.}
  \label{fig:timeliness}
  \vspace{-4mm}
\end{figure}

Timeliness (Fig~\ref{fig:accuracy}) is how soon a prefetcher can prefetch a cache block against its reference time.
The overprediction in \amc{} arises from the change in vertex/edges in the dynamic graph every iteration. 
As in BFS, the dynamic graph change marginally affect the overall graph traversal path for Road-CA input; therefore, the overprediction is lowest. In addition, \amc{} can benefit from a throttling mechanism to delay prefetches and gain the lost coverage from early prefetchers.

\subsubsection{Additional Off-chip Traffic}

\begin{figure}[h]
  \centering
  \includegraphics[width=1\columnwidth]{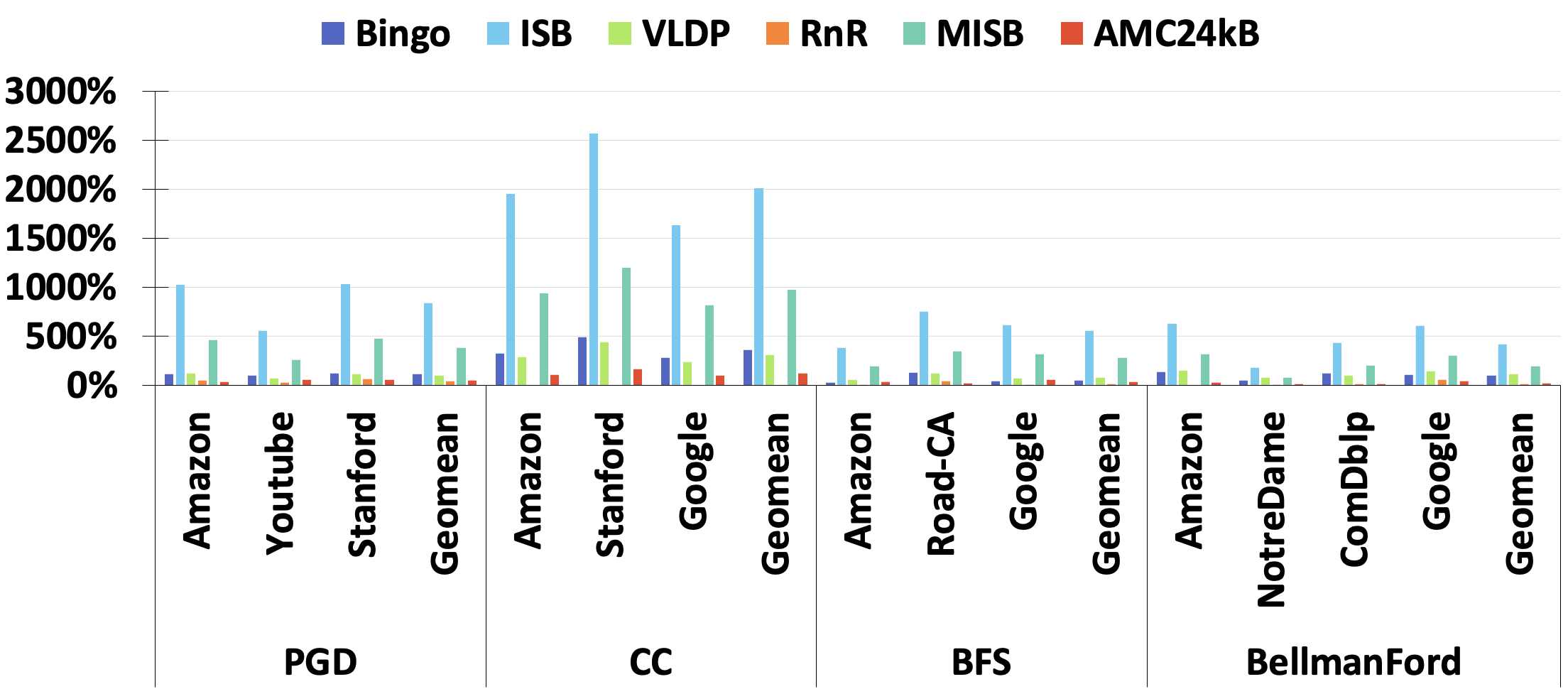}
  % \vspace*{-8mm}
  \caption{Additonal off-chip traffic.}
  \label{fig:offchipTraffic}
  \vspace{-4mm}
\end{figure}

\begin{figure}[h]
  \centering
  \includegraphics[width=\columnwidth]{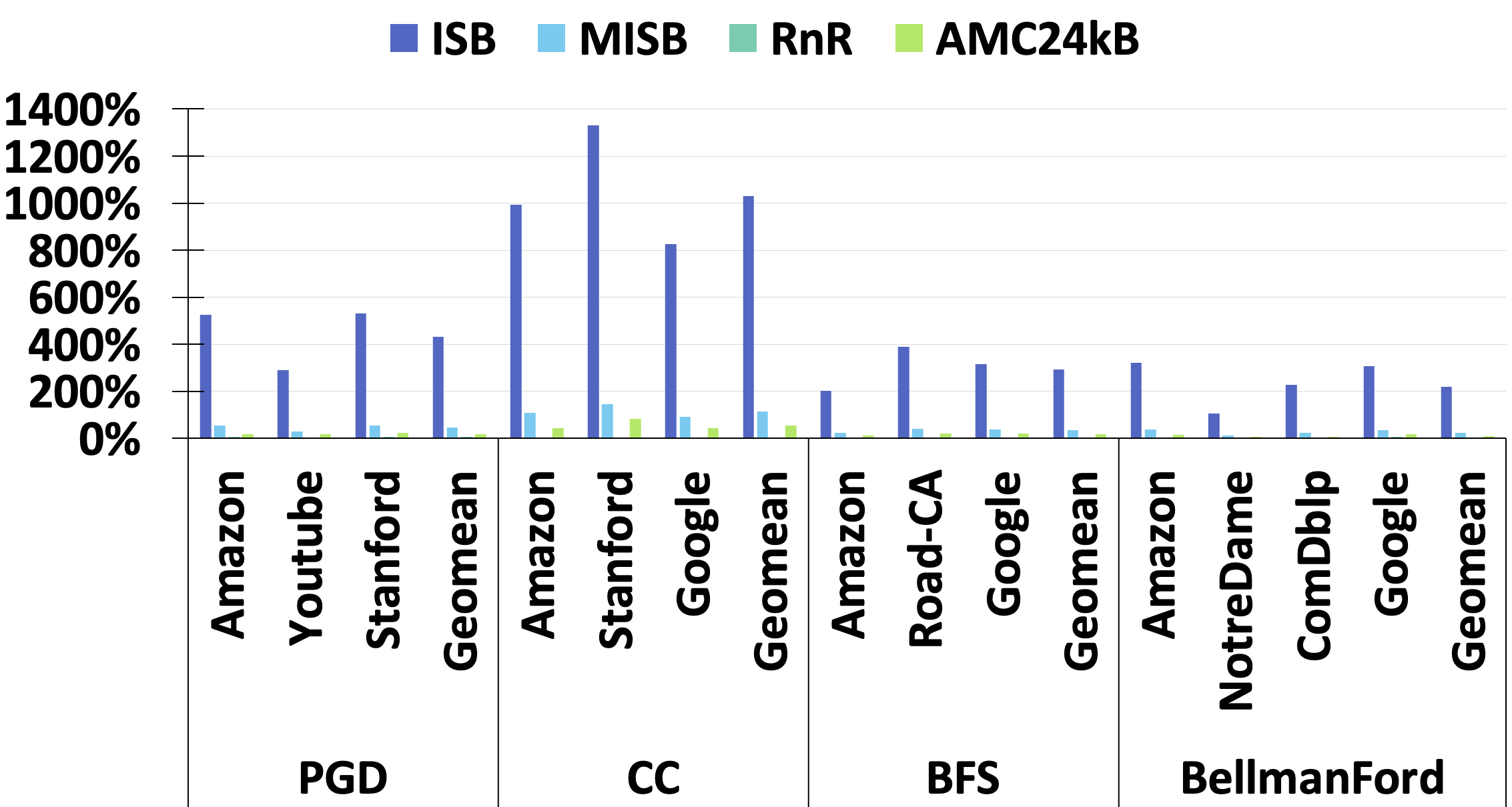}
  % \vspace*{-8mm}
  \caption{Off-chip metadata traffic.}
  \label{fig:MetaTraffic}
  \vspace{-2mm}
\end{figure}

The additional off-chip is $\frac{PrefDramAccess~ - ~DemandDramAccess}{DemandDramAccess}$ (Fig~\ref{fig:offchipTraffic}). PrefDramAccess is the number of main memory accesses with prefetchers. DemandDramAccess is the number of main memory accesses in the baseline.
On average, ISB and MISB issue 4$\times$  more prefetch than \amc{}. The high accuracy of \amc{} prefetcher and compressed metadata are the main reason for relatively low additional off-chip traffic. We break down additional off-chip traffic to determine the metadata traffic for prefetchers that use off-chip metadata storage (Fig~\ref{fig:MetaTraffic}). 
\amc{} metadata traffic on an average is 25\% compared to 493\% and 54\% for ISB and MISB respectively. The overall average additional off-chip traffic is 155\%, 958\%, 151\%, 458\%, and 56\% for Bingo, ISB, VLDP, MISB, and \amc{} respectively.

% \begin{figure}[ht]
%   \begin{minipage}[b]{0.48\textwidth}
%     \centering
%     \includegraphics[width=\textwidth]{Images/metadataTraffic.png}
%     \caption{Off-chip metadata traffic.}
%     \label{fig:MetaTraffic}
%   \end{minipage}
%   \hfill
%   \begin{minipage}[b]{0.48\textwidth}
%     \centering
%     \includegraphics[width=\textwidth]{Images/amc_power.png}
%     \caption{Energy comparison between baseline and \amc{} prefetcher.}
%     \label{fig:power}
%   \end{minipage}
% \end{figure}

\subsection{Energy Overhead}

\label{sec:energyOverhead}
\begin{figure}[h]
  \centering
  \includegraphics[width=0.8\columnwidth]{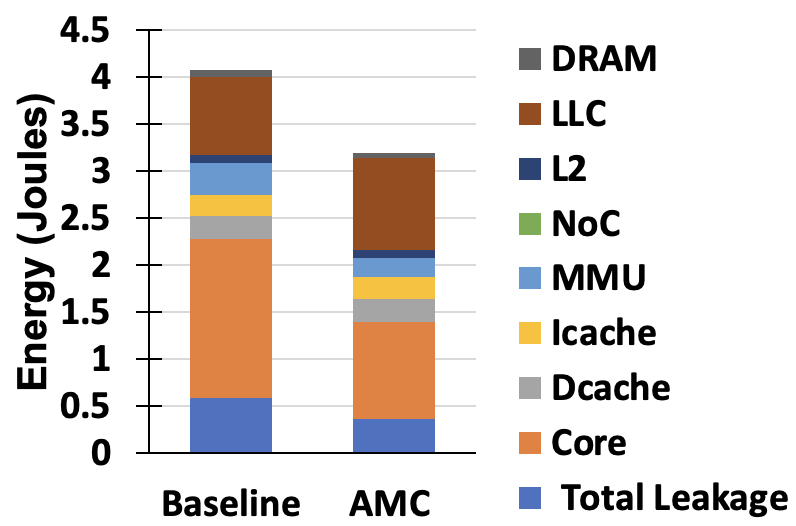}
  % \vspace*{-8mm}
  \caption{Energy comparison between baseline and \amc{} prefetcher.}
  \label{fig:power}
  \vspace{-2mm}
\end{figure}

Fig~\ref{fig:power} shows the energy breakdown and comparison between baseline and \amc{} prefetcher. 
The energy consumption for \amc{} prefetcher reduces in all the categories (core, cache, memory). \amc{} prefetcher consume on an average 1.28$\times$ less energy than baseline. This is chiefly because of reduced static energy consumption of core, caches, and DRAM because of reduced overall execution time.

\subsection{Hardware Overhead}
\label{sec:onChipStorage}

\amc{} requires a moderate amount of logic per core and set of architectural and internal registers (Section~\ref{sec:softwareSupport}) for compression and developing the correlation between L1 data accesses and L2 misses. The \amcCache{} requires about 29 kB for each core (78.3E$^{-3}$ mm$^{2}$) for each core. Base$\Delta$ compressor unit occupies about 13.3E$^{-3}$ mm$^{2}$ per core. The overall on-chip area is 0.2\% of the total on-chip area (46.19 mm$^{2}$).

\subsection{Storage Overhead}

\label{sec:offChipStorage}

\begin{figure}[h]
  \centering
  \includegraphics[width=1\columnwidth]{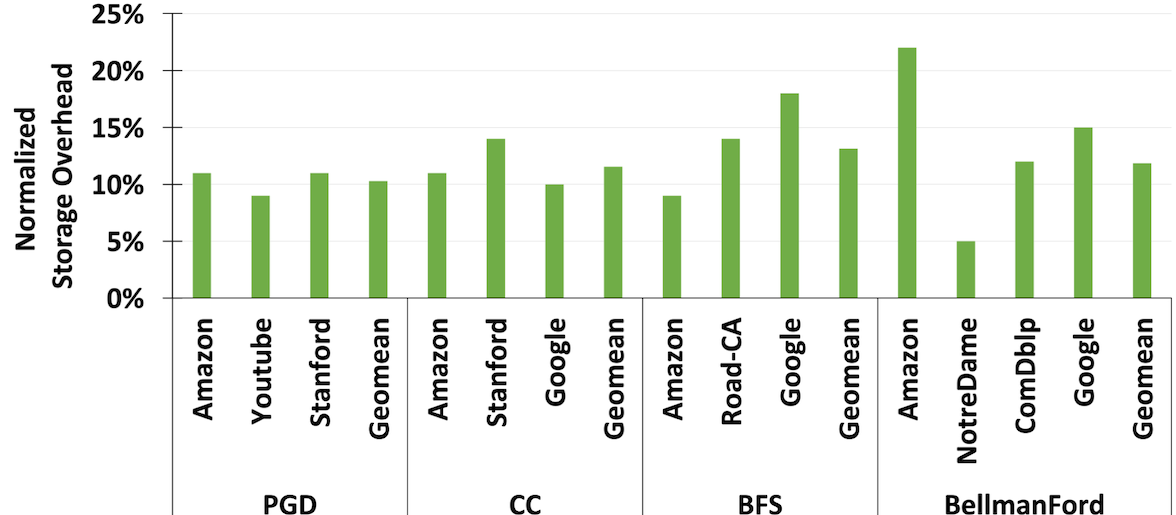}
  % \vspace*{-8mm}
  \caption{Off-chip metadata storage overhead.}
  \label{fig:offchipStorage}
  \vspace{-2mm}
\end{figure}

From Fig~\ref{fig:offchipStorage}, the off-chip storage is always below 25\% of the input size. If the kernel access pattern shows poor spatial locality, \amc{} needs to record a missing stream that can span multiple pages, which reduces the compression ratio of the overall missed stream.

\subsection{Miss Size Sensitivity}
\label{sec:missSizeStdy}
\begin{figure}[h]
  % \centering
  \includegraphics[width=.48\textwidth]{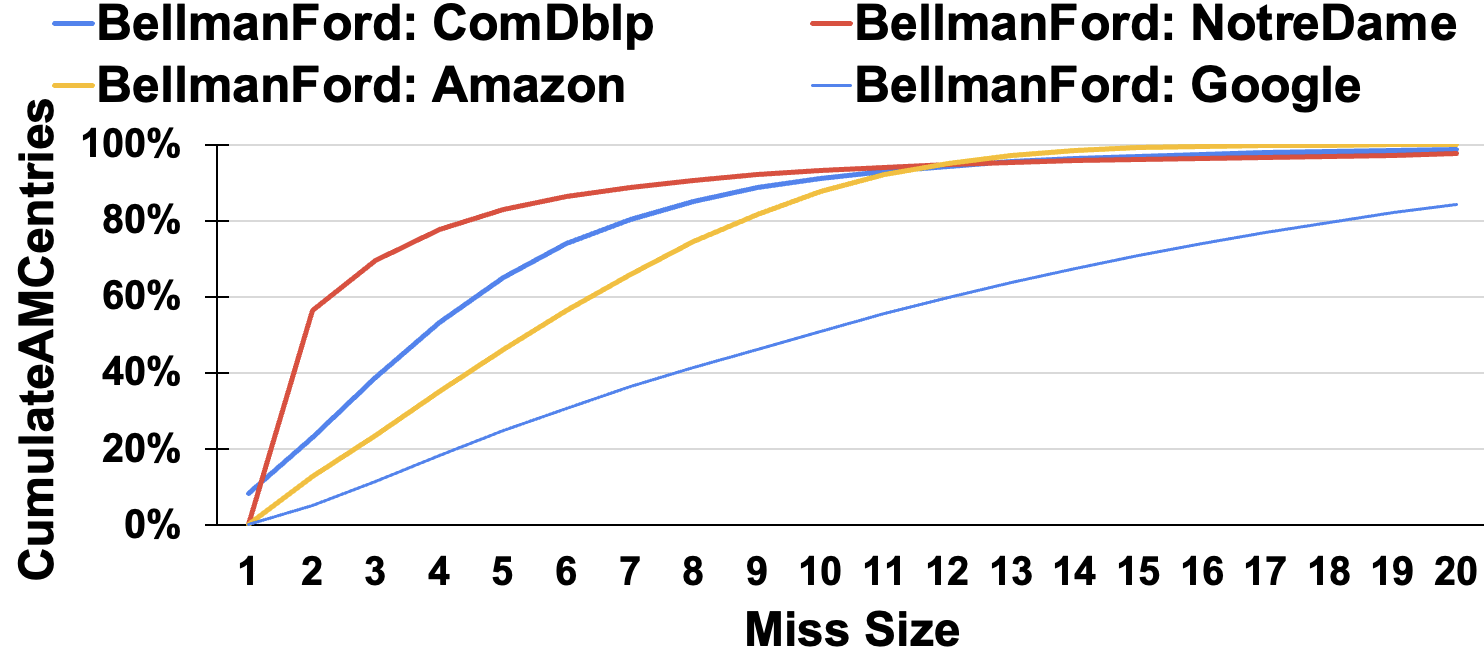}
  \includegraphics[width=.48\textwidth]{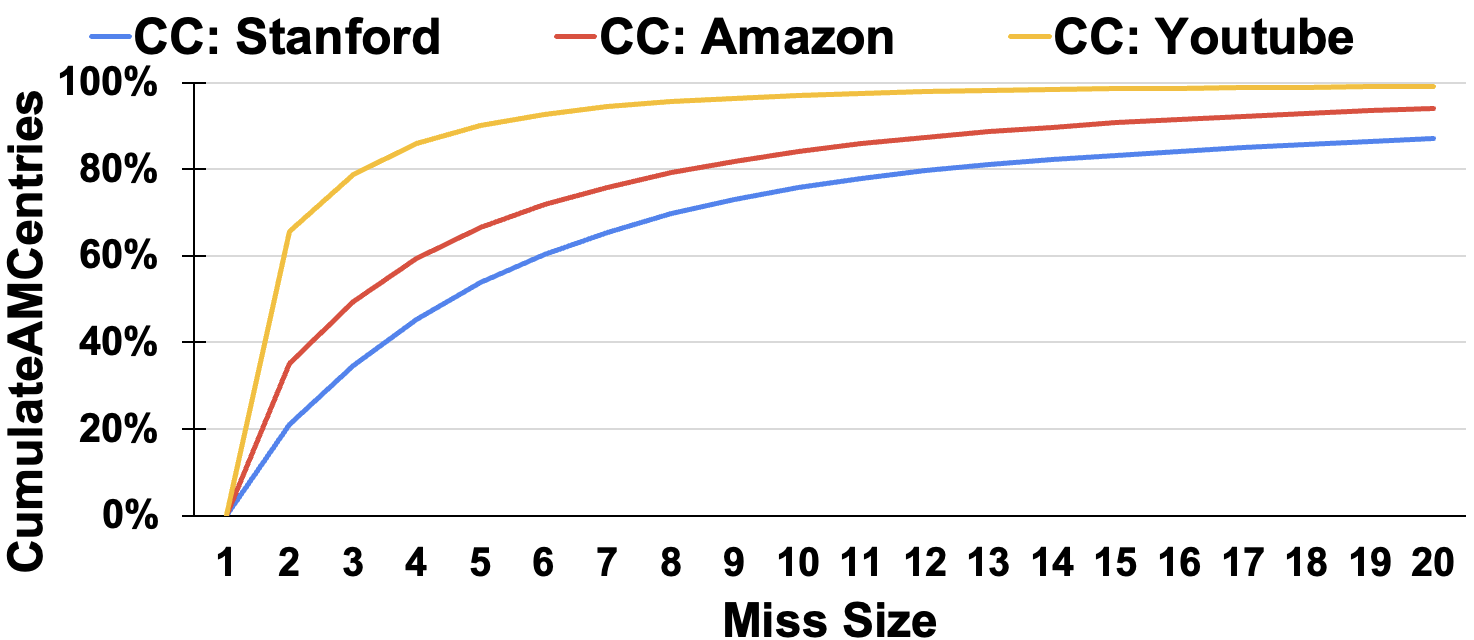}
  \includegraphics[width=.48\textwidth]{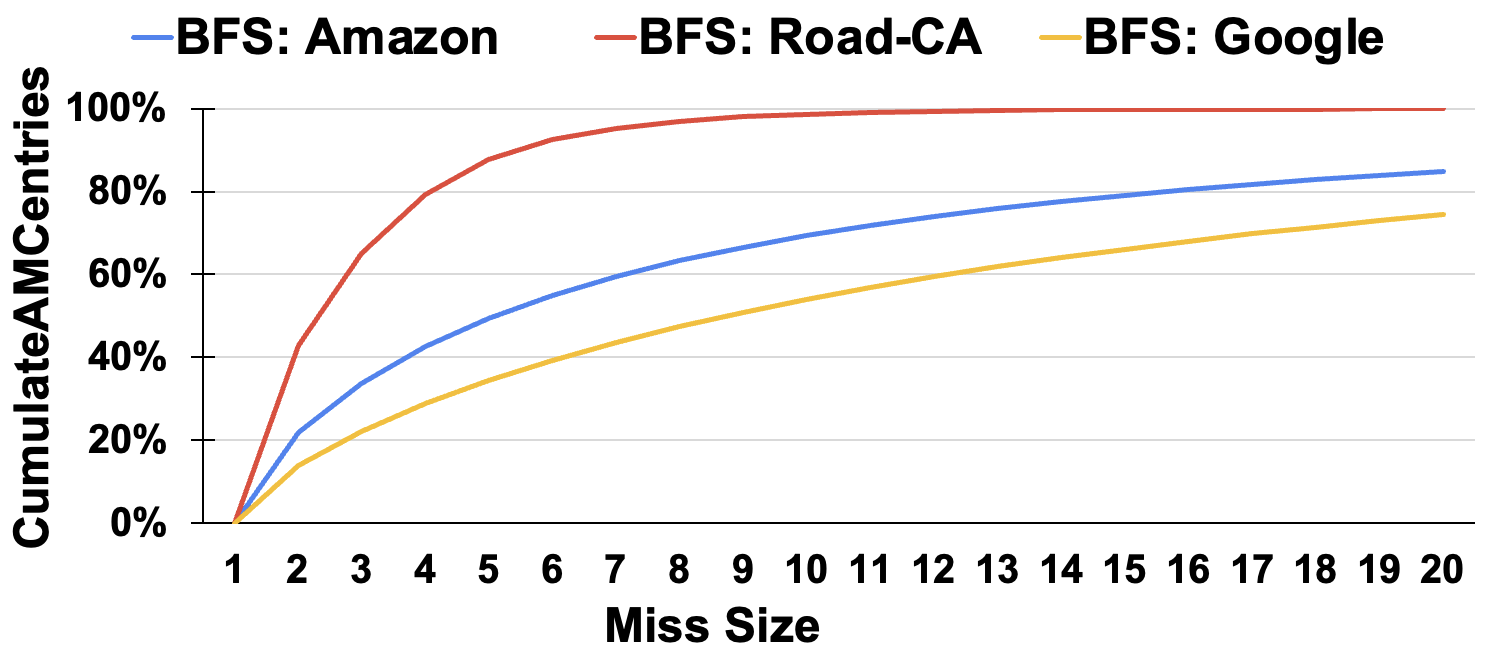}
  \includegraphics[width=.48\textwidth]{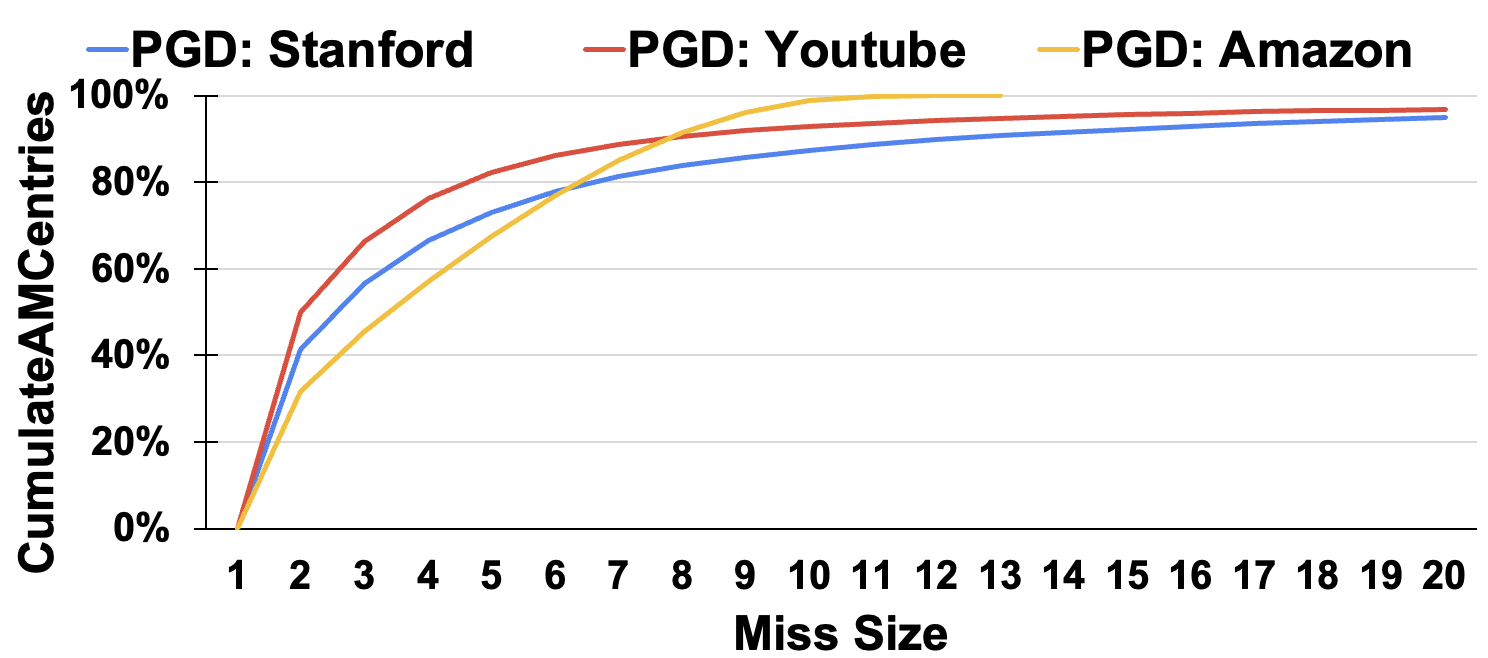}
  % \vspace*{-2mm}
  \caption{Miss size sensitivity.}
  \label{fig:missSensitivity}
  \vspace{-0.8ex}
\end{figure}

To select an appropriate miss stream size of an \amc{} entry. \amc{} assumes infinitely sized \amcCache{} with infinitely sized miss stream size of an \amc{} entry.
Fig~\ref{fig:missSensitivity} shows that the number of \amc{} entries with a miss size greater than 20 is less than 1\% for the evaluated kernel and input.
Another observation is 20 misses per \amc{} entry ensures evaluated kernel-input pair cover at least 74\% of entries. Increasing the number of misses beyond 20 does not yield much performance improvement that justifies the additional hardware overhead compared. Consequently, \amc{} records 20 misses per entry.

\section{Conclusion}
This work proposes a novel lightweight software-assisted \amc{} hardware prefetcher to improve prefetching accuracy and miss coverage for dynamic graph application. 
By allowing programmers to identify the target data structure, the proposed \amc{} prefetcher uses the ``many-to-many'' correlation style that adds contextual information to solve the aliasing problem and adapt to graph changes. \amc{} prefetcher stores compressed metadata both on-chip and off-chip, thus efficiently utilizing the memory bandwidth and space.

\bibliographystyle{plain}
\bibliography{refs}

\begin{thebibliography}{10}

\bibitem{ics}
Sam Ainsworth and Timothy~M Jones.
\newblock Graph prefetching using data structure knowledge.
\newblock In {\em Proceedings of the 2016 International Conference on Supercomputing}, pages 1--11, 2016.

\bibitem{ainsworth2017software}
Sam Ainsworth and Timothy~M Jones.
\newblock Software prefetching for indirect memory accesses.
\newblock In {\em 2017 IEEE/ACM International Symposium on Code Generation and Optimization (CGO)}, pages 305--317. IEEE, 2017.

\bibitem{event_driven}
Sam Ainsworth and Timothy~M Jones.
\newblock An event-triggered programmable prefetcher for irregular workloads.
\newblock {\em ACM Sigplan Notices}, 53(2):578--592, 2018.

\bibitem{compiler2003}
Hassan Al-Sukhni, Ian Bratt, and Daniel~A Connors.
\newblock Compiler-directed content-aware prefetching for dynamic data structures.
\newblock In {\em 2003 12th International Conference on Parallel Architectures and Compilation Techniques}, pages 91--100. IEEE, 2003.

\bibitem{bakhshalipour2018domino}
Mohammad Bakhshalipour, Pejman Lotfi-Kamran, and Hamid Sarbazi-Azad.
\newblock Domino temporal data prefetcher.
\newblock In {\em 2018 IEEE International Symposium on High Performance Computer Architecture (HPCA)}, pages 131--142. IEEE, 2018.

\bibitem{bingo}
Mohammad Bakhshalipour, Mehran Shakerinava, Pejman Lotfi-Kamran, and Hamid Sarbazi-Azad.
\newblock Bingo spatial data prefetcher.
\newblock In {\em 2019 IEEE International Symposium on High Performance Computer Architecture (HPCA)}, pages 399--411, 2019.

\bibitem{cacti}
Rajeev Balasubramonian, Andrew~B. Kahng, Naveen Muralimanohar, Ali Shafiee, and Vaishnav Srinivas.
\newblock Cacti 7: New tools for interconnect exploration in innovative off-chip memories.
\newblock {\em ACM Trans. Archit. Code Optim.}, 2017.

\bibitem{droplet}
Abanti Basak, Shuangchen Li, Xing Hu, Sang~Min Oh, Xinfeng Xie, Li~Zhao, Xiaowei Jiang, and Yuan Xie.
\newblock Analysis and optimization of the memory hierarchy for graph processing workloads.
\newblock In {\em 2019 IEEE International Symposium on High Performance Computer Architecture (HPCA)}, pages 373--386. IEEE, 2019.

\bibitem{hau}
Abanti Basak, Zheng Qu, Jilan Lin, Alaa~R Alameldeen, Zeshan Chishti, Yufei Ding, and Yuan Xie.
\newblock Improving streaming graph processing performance using input knowledge.
\newblock In {\em MICRO-54: 54th Annual IEEE/ACM International Symposium on Microarchitecture}, pages 1036--1050, 2021.

\bibitem{informedPref}
Mustafa Cavus, Resit Sendag, and Joshua~J. Yi.
\newblock Informed prefetching for indirect memory accesses.
\newblock {\em ACM Trans. Archit. Code Optim.}, 17(1), mar 2020.

\bibitem{champsim}
ChampSim.
\newblock Champsim simulator.
\newblock \url{https://github.com/ChampSim/ChampSim}, 2020.

\bibitem{telegraphcq}
Sirish Chandrasekaran, Owen Cooper, Amol Deshpande, Michael~J Franklin, Joseph~M Hellerstein, Wei Hong, Sailesh Krishnamurthy, Samuel~R Madden, Fred Reiss, and Mehul~A Shah.
\newblock Telegraphcq: continuous dataflow processing.
\newblock In {\em Proceedings of the 2003 ACM SIGMOD international conference on Management of data}, pages 668--668, 2003.

\bibitem{Kineograph}
Raymond Cheng, Ji~Hong, Aapo Kyrola, Youshan Miao, Xuetian Weng, Ming Wu, Fan Yang, Lidong Zhou, Feng Zhao, and Enhong Chen.
\newblock Kineograph: Taking the pulse of a fast-changing and connected world.
\newblock In {\em Proceedings of the 7th ACM European Conference on Computer Systems}, EuroSys '12, page 85–98, New York, NY, USA, 2012. Association for Computing Machinery.

\bibitem{spmd}
Frederica Darema, David~A George, V~Alan Norton, and Gregory~F Pfister.
\newblock A single-program-multiple-data computational model for epex/fortran.
\newblock {\em Parallel Computing}, 7(1):11--24, 1988.

\bibitem{sms}
Michael Ferdman, Stephen Somogyi, and Babak Falsafi.
\newblock Spatial memory streaming with rotated patterns.
\newblock {\em 1st JILP Data Prefetching Championship}, 29, 2009.

\bibitem{dynamicGraph}
Kathrin Hanauer, Monika Henzinger, and Christian Schulz.
\newblock Recent advances in fully dynamic graph algorithms.
\newblock {\em arXiv preprint arXiv:2102.11169}, 2021.

\bibitem{Ogdynamic}
Frank Harary and Gopal Gupta.
\newblock Dynamic graph models.
\newblock {\em Mathematical and Computer Modelling}, 25(7):79--87, 1997.

\bibitem{optMarkov}
Zhigang Hu, Margaret Martonosi, and Stefanos Kaxiras.
\newblock Tcp: Tag correlating prefetchers.
\newblock In {\em The Ninth International Symposium on High-Performance Computer Architecture, 2003. HPCA-9 2003. Proceedings.}, pages 317--326. IEEE, 2003.

\bibitem{stream}
Sorin Iacobovici, Lawrence Spracklen, Sudarshan Kadambi, Yuan Chou, and Santosh~G Abraham.
\newblock Effective stream-based and execution-based data prefetching.
\newblock In {\em Proceedings of the 18th annual international conference on Supercomputing}, pages 1--11, 2004.

\bibitem{corepara}
Intel.
\newblock Intel i7-6700 (skylake), 4.0 ghz (turbo boost), 14 nm.
\newblock \url{https://www.intel.com/content/www/us/en/processors/core/desktop- 6th-gen-core-family-datasheet-vol-1.html}, 2020.

\bibitem{composite_pref}
Intel.
\newblock Intel® 64 and ia-32 architectures optimization reference manual.
\newblock \url{file:///Users/explore/Downloads/248966-046A-software-optimization-manual.pdf}, 2023.

\bibitem{timeDyGraph}
Anand~Padmanabha Iyer, Li~Erran Li, Tathagata Das, and Ion Stoica.
\newblock Time-evolving graph processing at scale.
\newblock In {\em Proceedings of the fourth international workshop on graph data management experiences and systems}, pages 1--6, 2016.

\bibitem{isb}
Akanksha Jain and Calvin Lin.
\newblock Linearizing irregular memory accesses for improved correlated prefetching.
\newblock In {\em Proceedings of the 46th Annual IEEE/ACM International Symposium on Microarchitecture}, pages 247--259, 2013.

\bibitem{apt_get}
Saba Jamilan, Tanvir~Ahmed Khan, Grant Ayers, Baris Kasikci, and Heiner Litz.
\newblock Apt-get: Profile-guided timely software prefetching.
\newblock In {\em Proceedings of the Seventeenth European Conference on Computer Systems}, EuroSys '22, page 747–764, New York, NY, USA, 2022. Association for Computing Machinery.

\bibitem{branch}
Daniel~A Jim{\'e}nez and Calvin Lin.
\newblock Dynamic branch prediction with perceptrons.
\newblock In {\em Proceedings HPCA Seventh International Symposium on High-Performance Computer Architecture}, pages 197--206. IEEE, 2001.

\bibitem{markov}
Doug Joseph and Dirk Grunwald.
\newblock Prefetching using markov predictors.
\newblock In {\em Proceedings of the 24th annual international symposium on Computer architecture}, pages 252--263, 1997.

\bibitem{partition}
George Karypis and Vipin Kumar.
\newblock A fast and high quality multilevel scheme for partitioning irregular graphs.
\newblock {\em SIAM Journal on scientific Computing}, 20(1):359--392, 1998.

\bibitem{gretch}
Anirudh~Mohan Kaushik, Gennady Pekhimenko, and Hiren Patel.
\newblock Gretch: a hardware prefetcher for graph analytics.
\newblock {\em ACM Transactions on Architecture and Code Optimization (TACO)}, 18(2):1--25, 2021.

\bibitem{t2}
Sushant Kondguli and Michael Huang.
\newblock T2: A highly accurate and energy efficient stride prefetcher.
\newblock In {\em 2017 IEEE International Conference on Computer Design (ICCD)}, pages 373--376. IEEE, 2017.

\bibitem{dol}
Sushant Kondguli and Michael Huang.
\newblock Division of labor: A more effective approach to prefetching.
\newblock In {\em 2018 ACM/IEEE 45th Annual International Symposium on Computer Architecture (ISCA)}, pages 83--95. IEEE, 2018.

\bibitem{cam}
Anargyros Krikelis and Charles~C Weems.
\newblock Associative processing and processors.
\newblock {\em Computer}, 27(11):12--17, 1994.

\bibitem{graphone}
Pradeep Kumar and H~Howie Huang.
\newblock Graphone: A data store for real-time analytics on evolving graphs.
\newblock {\em ACM Transactions on Storage (TOS)}, 15(4):1--40, 2020.

\bibitem{datasets}
Jure Leskovec and Andrej Krevl.
\newblock {SNAP Datasets}: {Stanford} large network dataset collection.
\newblock \url{http://snap.stanford.edu/data}, June 2014.

\bibitem{mcpat}
Sheng Li, Jung~Ho Ahn, Richard~D. Strong, Jay~B. Brockman, Dean~M. Tullsen, and Norman~P. Jouppi.
\newblock {McPAT: An Integrated Power, Area, and Timing Modeling Framework for Multicore and Manycore Architectures}.
\newblock In {\em MICRO 42: Proceedings of the 42nd Annual IEEE/ACM International Symposium on Microarchitecture}, pages 469--480, 2009.

\bibitem{graphbolt}
Mugilan Mariappan and Keval Vora.
\newblock Graphbolt: Dependency-driven synchronous processing of streaming graphs.
\newblock In {\em Proceedings of the Fourteenth EuroSys Conference 2019}, pages 1--16, 2019.

\bibitem{dredge}
Andrew McCrabb, Eric Winsor, and Valeria Bertacco.
\newblock Dredge: Dynamic repartitioning during dynamic graph execution.
\newblock In {\em 2019 56th ACM/IEEE Design Automation Conference (DAC)}, pages 1--6. IEEE, 2019.

\bibitem{streamgraph}
Andrew McGregor.
\newblock Graph stream algorithms: a survey.
\newblock {\em ACM SIGMOD Record}, 43(1):9--20, 2014.

\bibitem{bestoffset}
Pierre Michaud.
\newblock Best-offset hardware prefetching.
\newblock In {\em 2016 IEEE International Symposium on High Performance Computer Architecture (HPCA)}, pages 469--480. IEEE, 2016.

\bibitem{ddr4power}
Micron.
\newblock Micron system power calculators, 2020.

\bibitem{software_prefetching}
Todd Mowry and Anoop Gupta.
\newblock Tolerating latency through software-controlled prefetching in shared-memory multiprocessors.
\newblock {\em Journal of parallel and Distributed Computing}, 12(2):87--106, 1991.

\bibitem{vr}
Ajeya Naithani, Sam Ainsworth, Timothy~M Jones, and Lieven Eeckhout.
\newblock Vector runahead.
\newblock In {\em 2021 ACM/IEEE 48th Annual International Symposium on Computer Architecture (ISCA)}, pages 195--208. IEEE, 2021.

\bibitem{dvr}
Ajeya Naithani, Jaime Roelandts, Sam Ainsworth, Timothy~M Jones, and Lieven Eeckhout.
\newblock Decoupled vector runahead.
\newblock In {\em Proceedings of the 56th Annual IEEE/ACM International Symposium on Microarchitecture}, pages 17--31, 2023.

\bibitem{ghb}
Kyle~J Nesbit and James~E Smith.
\newblock Data cache prefetching using a global history buffer.
\newblock In {\em 10th International Symposium on High Performance Computer Architecture (HPCA'04)}, pages 96--96. IEEE, 2004.

\bibitem{page1999pagerank}
Lawrence Page, Sergey Brin, Rajeev Motwani, and Terry Winograd.
\newblock The pagerank citation ranking: Bringing order to the web.
\newblock Technical report, Stanford InfoLab, 1999.

\bibitem{bouquet}
Samuel Pakalapati and Biswabandan Panda.
\newblock Bouquet of instruction pointers: Instruction pointer classifier-based spatial hardware prefetching.
\newblock In {\em 2020 ACM/IEEE 47th Annual International Symposium on Computer Architecture (ISCA)}, pages 118--131. IEEE, 2020.

\bibitem{baseDeltaCom}
Gennady Pekhimenko, Vivek Seshadri, Onur Mutlu, Phillip~B Gibbons, Michael~A Kozuch, and Todd~C Mowry.
\newblock Base-delta-immediate compression: Practical data compression for on-chip caches.
\newblock In {\em Proceedings of the 21st international conference on Parallel architectures and compilation techniques}, pages 377--388, 2012.

\bibitem{jetstream}
Shafiur Rahman, Mahbod Afarin, Nael Abu-Ghazaleh, and Rajiv Gupta.
\newblock Jetstream: Graph analytics on streaming data with event-driven hardware accelerator.
\newblock In {\em MICRO-54: 54th Annual IEEE/ACM International Symposium on Microarchitecture}, pages 1091--1105, 2021.

\bibitem{oltp}
Parthasarathy Ranganathan, Kourosh Gharachorloo, Sarita~V Adve, and Luiz~Andr{\'e} Barroso.
\newblock Performance of database workloads on shared-memory systems with out-of-order processors.
\newblock In {\em Proceedings of the eighth international conference on Architectural support for programming languages and operating systems}, pages 307--318, 1998.

\bibitem{twitter}
David Sayce.
\newblock The number of tweets per day in 2020, 2022.
\newblock https://www.dsayce.com/social-media/tweets-day/.

\bibitem{tage}
Andr{\'e} Seznec.
\newblock A 256 kbits l-tage branch predictor.
\newblock {\em Journal of Instruction-Level Parallelism (JILP) Special Issue: The Second Championship Branch Prediction Competition (CBP-2)}, 9:1--6, 2007.

\bibitem{vldp}
Manjunath Shevgoor, Sahil Koladiya, Rajeev Balasubramonian, Chris Wilkerson, Seth~H Pugsley, and Zeshan Chishti.
\newblock Efficiently prefetching complex address patterns.
\newblock In {\em Proceedings of the 48th International Symposium on Microarchitecture}, pages 141--152, 2015.

\bibitem{tornado}
Xiaogang Shi, Bin Cui, Yingxia Shao, and Yunhai Tong.
\newblock Tornado: A system for real-time iterative analysis over evolving data.
\newblock In {\em Proceedings of the 2016 International Conference on Management of Data}, pages 417--430, 2016.

\bibitem{ligra}
Julian Shun and Guy~E Blelloch.
\newblock Ligra: a lightweight graph processing framework for shared memory.
\newblock In {\em Proceedings of the 18th ACM SIGPLAN symposium on Principles and practice of parallel programming}, pages 135--146, 2013.

\bibitem{shun2016parallel}
Julian Shun, Farbod Roosta-Khorasani, Kimon Fountoulakis, and Michael~W Mahoney.
\newblock Parallel local graph clustering.
\newblock {\em arXiv preprint arXiv:1604.07515}, 2016.

\bibitem{synopsys}
Synopsys.
\newblock {Synopsys Standard Cell Libraries}.
\newblock \url{https://www.synopsys.com/dw/ipdir.php?ds=dwc_standard_cell}, 2019.
\newblock [Version P-2019.03, March 2019].

\bibitem{prodigy}
Nishil Talati, Kyle May, Armand Behroozi, Yichen Yang, Kuba Kaszyk, Christos Vasiladiotis, Tarunesh Verma, Lu~Li, Brandon Nguyen, Jiawen Sun, et~al.
\newblock Prodigy: Improving the memory latency of data-indirect irregular workloads using hardware-software co-design.
\newblock In {\em 2021 IEEE International Symposium on High-Performance Computer Architecture (HPCA)}, pages 654--667. IEEE, 2021.

\bibitem{tang2021dynamic}
Hao Tang, Guoshuai Zhao, Xuxiao Bu, and Xueming Qian.
\newblock Dynamic evolution of multi-graph based collaborative filtering for recommendation systems.
\newblock {\em Knowledge-Based Systems}, 228:107251, 2021.

\bibitem{dramcfg}
Micron Technology.
\newblock 8gb: x4, x8, x16 ddr4 sdram features.
\newblock \url{https://www.micron.com/-/media/client/global/documents/products/data-sheet/dram/ddr4/8gb_ddr4_sdram.pdf}.

\bibitem{vaziri2021controlling}
Pourya Vaziri and Keval Vora.
\newblock Controlling memory footprint of stateful streaming graph processing.
\newblock In {\em 2021 USENIX Annual Technical Conference (USENIX ATC 21)}, pages 269--283, 2021.

\bibitem{metasys}
Nandita Vijaykumar, Ataberk Olgun, Konstantinos Kanellopoulos, F~Nisa Bostanci, Hasan Hassan, Mehrshad Lotfi, Phillip~B Gibbons, and Onur Mutlu.
\newblock Metasys: A practical open-source metadata management system to implement and evaluate cross-layer optimizations.
\newblock {\em ACM Transactions on Architecture and Code Optimization (TACO)}, 19(2):1--29, 2022.

\bibitem{vora2017kickstarter}
Keval Vora, Rajiv Gupta, and Guoqing Xu.
\newblock Kickstarter: Fast and accurate computations on streaming graphs via trimmed approximations.
\newblock In {\em Proceedings of the twenty-second international conference on architectural support for programming languages and operating systems}, pages 237--251, 2017.

\bibitem{grasu}
Qinggang Wang, Long Zheng, Yu~Huang, Pengcheng Yao, Chuangyi Gui, Xiaofei Liao, Hai Jin, Wenbin Jiang, and Fubing Mao.
\newblock Grasu: A fast graph update library for fpga-based dynamic graph processing.
\newblock In {\em The 2021 ACM/SIGDA International Symposium on Field-Programmable Gate Arrays}, pages 149--159, 2021.

\bibitem{wen2017fog}
Zhenyu Wen, Renyu Yang, Peter Garraghan, Tao Lin, Jie Xu, and Michael Rovatsos.
\newblock Fog orchestration for internet of things services.
\newblock {\em IEEE Internet Computing}, 21(2):16--24, 2017.

\bibitem{stms}
Thomas~F Wenisch, Michael Ferdman, Anastasia Ailamaki, Babak Falsafi, and Andreas Moshovos.
\newblock Practical off-chip meta-data for temporal memory streaming.
\newblock In {\em 2009 IEEE 15th International Symposium on High Performance Computer Architecture}, pages 79--90. IEEE, 2009.

\bibitem{thesisMisb}
Hao Wu et~al.
\newblock {\em Practical irregular prefetching}.
\newblock PhD thesis, The University of Texas at Austin, 2020.

\bibitem{triage}
Hao Wu, Krishnendra Nathella, Joseph Pusdesris, Dam Sunwoo, Akanksha Jain, and Calvin Lin.
\newblock Temporal prefetching without the off-chip metadata.
\newblock In {\em Proceedings of the 52nd Annual IEEE/ACM International Symposium on Microarchitecture}, pages 996--1008, 2019.

\bibitem{misb}
Hao Wu, Krishnendra Nathella, Dam Sunwoo, Akanksha Jain, and Calvin Lin.
\newblock Efficient metadata management for irregular data prefetching.
\newblock In {\em 2019 ACM/IEEE 46th Annual International Symposium on Computer Architecture (ISCA)}, pages 1--13. IEEE, 2019.

\bibitem{rnr}
Chao Zhang, Yuan Zeng, John Shalf, and Xiaochen Guo.
\newblock Rnr: A software-assisted record-and-replay hardware prefetcher.
\newblock In {\em 2020 53rd Annual IEEE/ACM International Symposium on Microarchitecture (MICRO)}, pages 609--621. IEEE, 2020.

\bibitem{minnow}
Dan Zhang, Xiaoyu Ma, Michael Thomson, and Derek Chiou.
\newblock Minnow: Lightweight offload engines for worklist management and worklist-directed prefetching.
\newblock {\em ACM SIGPLAN Notices}, 53(2):593--607, 2018.

\bibitem{tdgraph}
Jin Zhao, Yun Yang, Yu~Zhang, Xiaofei Liao, Lin Gu, Ligang He, Bingsheng He, Hai Jin, Haikun Liu, Xinyu Jiang, et~al.
\newblock Tdgraph: a topology-driven accelerator for high-performance streaming graph processing.
\newblock In {\em Proceedings of the 49th Annual International Symposium on Computer Architecture}, pages 116--129, 2022.

\end{thebibliography}

\end{document}